\documentclass[10pt,reprint,notitlepage,a4paper,superscriptaddress,floatfix,longbibliography,amsmath,amssymb,aps,prl]{revtex4-2}
\usepackage[T1]{fontenc}
\usepackage{babel}
\usepackage{amsmath}
\usepackage{amssymb}
\usepackage{graphicx}
\usepackage{bm}
\usepackage{ulem}
\usepackage{amsmath}
\usepackage{graphicx}
\usepackage[unicode=true,pdfusetitle,
urlcolor=blue,bookmarks=true,bookmarksnumbered=false,bookmarksopen=false,
 breaklinks=false,pdfborder={0 0 1},backref=false,colorlinks=true]{hyperref}
\usepackage{comment}
\usepackage[percent]{overpic}
\usepackage{xr-hyper}
\usepackage{subfiles}
\usepackage{float} 

\DeclareMathOperator{\Extr}{Extr}
\newcommand{\figpanel}[1]{(\textbf{\lowercase{#1}})}
\usepackage{xcolor}
\definecolor{blue(ncs)}{rgb}{0.0, 0.53, 0.74}

\definecolor{truc(ncs)}{rgb}{0.5, 0.53, 0.74}
\definecolor{violet(ncs)}{rgb}{0.58, 0.0, 0.83}

\definecolor{red(ncs)}{rgb}{0.90, 0.40, 0.40}

\definecolor{mygreen}{RGB}{26, 148, 49}

\newcommand{\xip}{{\hat{\xi}}}
\newcommand{\meanv}[1]{\left\langle#1\right\rangle}
\newcommand{\set}[1]{{\left\{#1\right\}}} 
\usepackage{appendix}
\usepackage{silence}

\makeatletter

\newcommand{\apptocfile}{atoc}

\let\apptoc@orig@appendix\appendix
\renewcommand{\appendix}{%
  \apptoc@orig@appendix
  \let\apptoc@orig@addtocontents\addtocontents
  \long\def\addtocontents##1##2{%
    \def\apptoc@ext{##1}%
    \def\apptoc@toc{toc}%
    \ifx\apptoc@ext\apptoc@toc
      \apptoc@orig@addtocontents{\apptocfile}{##2}%
    \else
      \apptoc@orig@addtocontents{##1}{##2}%
    \fi
  }%
}

\newcommand{\appendixtableofcontents}{%
  \begingroup
    \setcounter{tocdepth}{3}%
    \phantomsection
    \let\addcontentsline\@gobblethree
    \section*{Appendix\\Supplemental Material}%
    \pdfbookmark[1]{Appendices}{apxcontents}%
    \@starttoc{\apptocfile}%
  \endgroup
}

\begin{document}

\preprint{APS/123-QED}

\title{A solvable model for unsupervised federated learning}

\author{Giovanni Catania}
\affiliation{Institute for Cross-disciplinary Physics and Complex Systems IFISC (CSIC-UIB),
Campus Universitat Illes Balears, 07122 Palma de Mallorca, Spain.}
\affiliation{Departamento de Física Teórica, Universidad Complutense de Madrid,
28040 Madrid, Spain.}
\author{Aur\'{e}lien Decelle}
\affiliation{Escuela Técnica Superior de Ingenieros Industriales, Universidad Politécnica de Madrid, Calle de José Gutiérrez Abascal 2, Madrid 28006,
Spain.}
\affiliation{GISC - Grupo Interdisciplinar de Sistemas Complejos 28040 Madrid, Spain.}
\author{Gianluca Manzan}
\email{gianluca.manzan@inria.fr}
\affiliation{Inria Saclay - Tau team, B\^at 660 Universit\'e Paris-Saclay, Orsay Cedex 91405}
\affiliation{LISN, Tau team, B\^at 660 Universit\'e Paris-Saclay, Orsay Cedex 91405}
\affiliation{Departamento de Física Teórica, Universidad Complutense de Madrid,
28040 Madrid, Spain.}
\affiliation{Department of Mathematics, University of Bologna, Piazza di Porta San Donato 5, 40126, Bologna (BO), Italy.}
\author{Beatriz Seoane}
\affiliation{Departamento de Física Teórica  \& IPARCOS, Universidad Complutense de Madrid, 28040 Madrid, Spain.}
\affiliation{GISC - Grupo Interdisciplinar de Sistemas Complejos 28040 Madrid, Spain.}
\author{Daniele Tantari}
\affiliation{Department of Mathematics, University of Bologna, Piazza di Porta San Donato 5, 40126, Bologna (BO), Italy.}

\date{\today}

\begin{abstract}
We introduce a theoretical framework for analyzing federated learning in a generative setting through a teacher-multiple interacting students scenario, in which each student receives a distinct realization of the data, either through a different noise corruption or by accessing a different subset, possibly of varying size. Using theoretical tools in equilibrium disordered system, we analytically show that interactions among students systematically enhance learning performance: highly noisy students require fewer samples to recover the underlying pattern, while low-noise students achieve a larger overlap with the ground-truth signal. We derive the optimal Bayesian conditions for teacher recovery as functions of the sample complexity, noise level, and interaction strength, and validate these predictions through numerical simulations. The resulting dynamics can be mapped onto equilibrium sampling in a Restricted Boltzmann Machine with a structured hidden layer, providing a principled theoretical understanding of how interactions improve distributed generative modeling. 

\end{abstract}

\maketitle
\textbf{Introduction---}Federated learning (FL) is a decentralized machine-learning framework in which multiple clients collaboratively train a global model while keeping their local data private~\cite{mcmahan2017communication, kairouz2021advances}. It is often adopted to distribute computational and communication resources or to balance training workloads across participants. In many settings, however, FL is not merely a matter of efficiency but a necessity, particularly when training relies on sensitive data that cannot be centralized or shared, as in large-scale communication systems or healthcare applications~\cite{kairouz2021advances,li2025challengespitfallsrecommendationsopportunities,Teo2024-af}.
Unlike centralized training, FL allows each client to compute local updates using its own dataset and to transmit only model information—such as gradients or parameter updates—to a central server~\cite{NIPS2012_6aca9700}. The server then aggregates these contributions, most commonly through federated averaging, to iteratively refine a global model. This procedure preserves data locality while still leveraging the statistical information distributed across the participating clients~\cite{rafi_fairness_privacy_FL}.

Despite its widespread practical adoption and a vast literature proposing heuristic solutions for client selection, aggregation, and communication efficiency~\cite{ayeelyan2024federated}, a detailed theoretical understanding of how the different components of FL influence performance remains limited~\cite{kairouz2021advances}. In this perspective, statistical physics provides a natural framework to address this gap by recasting FL within the teacher-student paradigm~\cite{gardner_three_1989,sompolinskylearningfromexamples, zdeborova2016statistical}: a teacher generates and distributes data, while multiple students (representing different clients) attempt to infer the underlying rule from their own data realizations while interacting through a shared learning protocol.
Interestingly, in recent years, such teacher-student approach has already demonstrated that coupling multiple learners can significantly accelerate training and optimization in a variety of computer science problems. In particular, studies of binary classification with the perceptron~\cite{baldassi_unreasonable_2016,baldassi_subdominant_dense_clusters,copycat_perceptron_PRE} and of the graph coloring problem~\cite{MC_PRX_coloring_RSA} have shown that replicating multiple students and allowing them to interact (rather than relying on a single-student setup) can substantially accelerate convergence toward solutions with improved generalization properties, while facilitating escape from metastable states during training. 
It is worth mentioning however, that this behavior seems to be related to the inference-related task and do not generalize to purely optimization process as studied in~\cite{louise_bicoloring_replicated}. Nevertheless, this protocol was also shown to be more efficient in deep learning context~\cite{elastic_SGD_lecun,Chaudhari_2019} on classification tasks. \\ When translated to the FL setting, these results suggest that FL should not be viewed merely as a practical constraint imposed by data locality or privacy considerations. Rather, it may constitute a more powerful and efficient training paradigm for large neural network models, potentially enabling faster convergence even when individual clients have access to limited or low-quality data.

To bridge previous results on coupled learning systems with realistic FL settings, we extend the theory to generative learning tasks in which multiple students, each with access only to a local dataset, infer the interaction parameters of a probabilistic teacher model. These datasets correspond to distinct realizations of the same underlying process and may differ across students in size or noise level. The goal is to learn generative models capable of reproducing the teacher’s distribution. Using the replica approach, we show that interactions among students are generically beneficial. Cooperative learning reduces the sample complexity required for recovery in high-noise regimes and enhances the overlap with the true signal, compared to isolated learning. We derive the recovery equations as functions of dataset size, noise level, coupling strength, and the number of learners, and validate our predictions through numerical simulations.
Finally, we identify a practical learning scheme in which the coupled-student dynamics correspond to equilibrium configurations of a Restricted Boltzmann Machine (RBM) with a structured hidden layer. Within this framework, we provide Bayes-optimal predictions for the conditions under which the students recover the teacher, quantifying the roles of data heterogeneity, noise amplitude, and inter-student coupling.

\textbf{Definition of the model and training tasks---} We assume that the training data are generated by a \textit{teacher} described by a Mattis model with a single fixed pattern $\hat{\boldsymbol{\xi}}$ with binary components, $\hat{\xi}_i \in \pm 1$ for $i = 1,\ldots,N$. The teacher produces a set of discrete configurations $\bm{\mathcal S} = \set{\bm{s}^\mu}_{\mu=1}^{M}$, with $s_i^\mu \in \set{\pm 1}$, sampled from the Boltzmann distribution associated with the Hamiltonian $\mathcal{H} = -N^{-1}\sum_{i<j} \hat{\xi}_i \hat{\xi}_j s_i s_j$ at inverse temperature $\hat{\beta}$. We then distribute the data among $y$ students, each of which aims to infer the underlying pattern $\hat{\bm{\xi}}$ used to generate the data. Before assigning the data to the students, we apply two different transformations to the dataset $\bm{\mathcal{S}}$ in order to generate student-specific versions $\bm{\mathcal J}^{u}=\set{\eta^{\mu_u}}_{\mu_u=1}^{M^u}$ where the index $u$ labels the student and $M^u$ the number of samples it receives. These transformations allow us to investigate different scenarios relevant for FL:

\textbf{(i) Student-dependent noise (SDN).}
In this setting, each student receives a different corrupted version of the same dataset $\bm{\mathcal S}$,
consisting of $M$ samples, with corruption induced by student‑dependent noise.
For student $u$, the example $\bm{s}^\mu$ is modified by a random vector $\bm{\chi}^{u\mu} \in \{\pm1\}^N$, so that the observed configuration is given by the element-wise product $\bm{\eta}^{\mu u}=\bm{s}^\mu \circ \bm{\chi}^{\mu u}$.
The components of $\bm{\chi}^{u\mu}$ are i.i.d. drawn according to $
        P(\chi_{i}^{u\mu})=\frac{1\!+\!r^{u}}{2}\delta(\chi_{i}^{u\mu}\!-\!1)\!+\!\frac{1\!-\!r^{u}}{2}\delta(\chi_{i}^{u\mu}\!+\!1)$,
where the parameter $r^u \in [0,1]$ controls the corruption level for student $u$ \cite{centonze2024statistical, agliari2024hebbian}. Noise realizations are independent across students, so that each student observes a different corrupted versions of the same teacher-generated dataset. Consequently, all datasets have identical cardinality, $M^u=M$.
In the limiting case $r^u = 1$, data are uncorrupted, whereas for $r^u = 0$ they are fully randomized, corresponding to flipping with probability $p = 1/2$.  
This construction allows for heterogeneous data quality across students. We denote by $\bm r$ the vector with components $\set{r^u}_{u=1}^{y}$, which encodes the corruption level associated with each student.

\textbf{(ii) Student-dependent dataset (SDD)}.  
    In this setting, the dataset of $M$ samples is partitioned among the students. Each student $u$ receives an independent subset of $M^u$ patterns drawn from the Boltzmann distribution associated with the teacher’s Mattis model, subject to the global constraint
\(\sum_u M^u = M\). We assume that each student has access to an extensive dataset, namely $\alpha^{u}  \!=  \!M^{u}/N \!\sim  \!O(1)$ in the thermodynamic limit. We collect these sample complexities into a vector $\bm{\alpha}=\left\{\alpha^u\right\}_{u=1, \ldots,y} $. 
The $y$ students are individually modeled by Mattis models with Ising spins and student-dependent binary weights $\bm\xi^u$. Within a Bayesian framework, when a student $u$ is considered in isolation, the posterior distribution $P(\bm{\xi^u}|\bm{\mathcal{J}}^u)$ is described by a Hopfield model \cite{hopfield1982neural} whose patterns are given by the samples provided by the teacher \cite{barra2018phase, alemanno2023hopfield, theriault2025modeling}. The inference task then reduces to characterizing the typical spin configurations, now $\set{\xi_i^u}_{i=1,\cdots,N}$, of this model.\\
To analyze the thermodynamics of the FL framework, we include a ferromagnetic interaction of strength $\gamma$ that promotes alignment among the students’ weights. This interaction can be interpreted as a prior favoring their mutual alignment, while each student retains access exclusively to its own local dataset. The posterior of the interacting system is then governed by the corresponding joint probability distribution:
\begin{equation}
P(\bm{\xi}|\bm{\mathcal{J}})\! =\!\frac{1}{{Z(\bm{\mathcal{J}})}}\exp\left(\!\frac{\beta}{N}\!\sum_{\substack{u,\mu_u \\
i<j}}\eta_i^{\mu_u} \eta_{j}^{\mu_u}\xi_{i}^{u}\xi_{j}^{u}\!+\!\frac{\gamma}{y}\sum_{\substack{u<v,\\i}}\xi_{i}^{u}\xi_{i}^{v}\right) 
\label{eq:probaBoltz}
\end{equation}
\noindent where $\bm{\mathcal J} \!=\!\set{\bm{\mathcal J}^{u}}_{u=1}^y$ and $\bm{\xi} \!=\!\set{\bm{\xi}^{u}}_{u=1}^y$ denote, respectively, the joint training dataset and the set of patterns of the $y$ students. Indices $u,v$ label the students, $i,j$ label the sites, and $\mu,\nu$ label the examples, while $\beta$ denotes the inverse temperature common to all students.
In this framework, the coupled system represents an ensemble of models that leverage mutual interactions to improve statistical inference. In the presence of corrupted data, cooperative couplings reinforce the shared signal, thereby enhancing the effective signal-to-noise ratio and promoting the recovery of the latent structure.
Through these interactions, information is pooled across learners: 
the ferromagnetic coupling biases the collective dynamics toward coherent configurations, effectively sharing the information among the students.

\textbf{Thermodynamics and replica computation of the free-energy ---} The evaluation of the equilibrium properties of the multi-student system relies on the analysis of the properties of the quenched free energy in the thermodynamic limit. 
Throughout this work, we focus on a noisy and non‑trivial inference regime,
corresponding to the teacher operating in the paramagnetic phase, \(\hat{\beta}<1\).
 The average behavior of the coupled system can be naturally analyzed using the replica trick \cite{mezard1987spin,charbonneau2023spin}, which we employ to compute its free energy:
\begin{eqnarray}
\label{eq: limiting free energy}
-\beta f(\Lambda)=\lim_{N\to\infty}\frac{1}{N}\left[\ln Z\right]^{\bm{\mathcal{J}}}\,,
\end{eqnarray}
where $\left[\cdots\right]^{\bm{\mathcal{J}}}$ denotes the average over the disorder $\bm{\mathcal{J}}$, and $\Lambda=\{y,\gamma,\beta,\hat{\beta},\bm\alpha,\bm{r}\}$ collects the control parameters of the model. Within this framework, SDN is characterized by $\bm\alpha\!=\!\alpha\bm 1,\bm{r}\!\neq\!\bm 1$, whereas SDD has $\bm{r}\!=\!\bm 1$. In both cases, the {\it working alone} (WA) limit is correctly retrieved by fixing $\gamma=0$.
The explicit expression of $\beta f(\Lambda)$ in the two scenarios is reported in the Supplemental Material (SM).

As is standard in the analysis of disordered systems, we introduce a set of order parameters to characterize the problem. In the present setting, these include the magnetization of each student with respect to the teacher, computed for a given independent sample from the posterior (often referred to  as a {\it replica}) $a$, denoted $m_u^a$. This measures the degree of alignment of $\bm \xi^u$ with the teacher’s pattern $\bm{\hat \xi}$. We also introduce the pairwise overlap between the weights of students $u$ and $v$, computed across replicas $a$ and $b$, which quantifies the similarity between the patterns inferred by different students, as well as their variability across independent posterior samples:
\begin{equation}
    m^a_u = \frac{1}{N}\sum_{i=1}^N\hat\xi_i \xi_i^{au} \,, \qquad q^{ab}_{uv} = \frac{1}{N} \sum_{i=1}^N \xi_i^{au} \xi_i^{bv}\,.
\end{equation}
As in \cite{copycat_perceptron_PRE}, we adopt a suitable Replica Symmetric (RS) ansatz, whose structure depends on the source of heterogeneity considered in the data.
We assume a generic structure for the overlap matrix: in the diagonal replica blocks ($a\!=\!b$), we take a uniform overlap between different students $u\neq v$, namely $q^{aa}_{uv}\!=\!\delta_{uv}\!+\!t(1-\delta_{uv})$. In the off-diagonal replica blocks ($a\!\neq \!b$), we assign a possibly different value $q\neq t$ for $u\neq v$ while $q^{ab}_{uu}\!=\!q_u$, reflecting the students can be exposed to the different dataset realizations. We further assume replica-symmetric magnetizations, $m^a_u\!=\!m^u$.
Further details on the analytic derivations are given in Sec. B for the SDN scenario (resp. C for the SDD) of the SM.
In all cases, we obtain a closed-form expression for $\beta f(\Lambda)$, which allows us to determine the phase diagrams as a function of the hyper-parameters $\Lambda$. 
The phase diagrams describe the equilibrium properties of the system through $\bm m=(m_1,...,m_y)$ and $\bm q=(q_1,...,q_y)$. The system exhibits three distinct phases: (i) a paramagnetic (P) phase, characterized by $\bm m = 0$; (ii) a signal-retrieval (sR) phase, where $\bm m > \bm 0$ and $\bm q > \bm 0$; and (iii) a spin-glass (SG) phase, with $\bm m = \bm 0$ and $\bm q > \bm 0$. 
Inference is only possible in the sR phase which therefore quantifies how the different parameters affect successful recovery.

\begin{figure}[t!]
\begin{overpic}[width=\columnwidth]{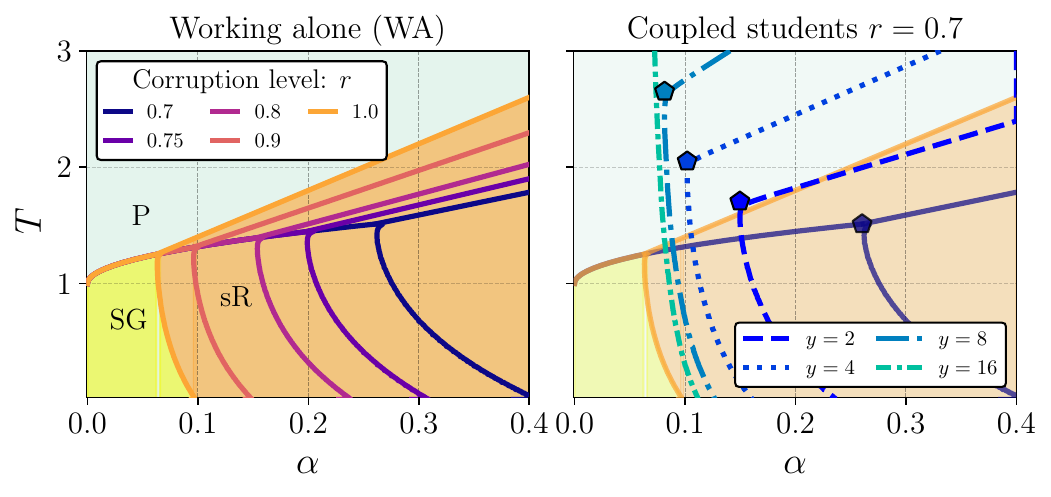}
\put(7,45.5){{\figpanel{a}}}
\put(53,45.5){{\figpanel{b}}}
\end{overpic}
    \caption{Phase diagrams for the recovery of the teacher pattern from samples generated at $\hat{\beta}\!=\!0.8$, using data corrupted at different noise levels parametrized by $r$.
\textbf{(a)} Working-alone (WA) scenario, in which students do not interact. The paramagnetic (P), spin-glass (SG), and signal-retrieval (sR) phases are indicated, with background colors corresponding to the uncorrupted data limit ($r\!=\!1$) analyzed in \cite{alemanno2023hopfield,manzan2025effect}. As the noise level increases (i.e., as $r$ decreases), the sR phase progressively shifts toward larger values of $\alpha$.
\textbf{(b)} Homogeneous-corruption SDN scenario with corruption level $r\!=\!0.7$. Different color lines indicate the retrieval region associated with varying numbers of students $y$ cooperating with $\gamma\!=\!1$. The shaded area represents the three phases for the WA at $r=0.7$.}
    \label{Fig:PD_WA}
\end{figure}

\textbf{Results ---} In the following, we assess the performance of the different FL strategies relative to the WA limit, where each student optimizes its model using only its local dataset. To this end, Fig.~\ref{Fig:PD_WA}--(a) displays the phase diagram in the $(\alpha,T)$ plane for the WA regime, in which students are provided with datasets exhibiting varying corruption levels $r$.
When noise dominates ($\hat{\beta}<1$) students can condense the information contained in the dataset once a threshold in data availability is reached. The triple point $P_\mathrm{c}=(\alpha_\mathrm{c},T_\mathrm{c})$ at the intersection of the three phases, marks the minimal recovery threshold with $\alpha_\mathrm{c}=\left(\frac{1-\hat{\beta}}{\hat{\beta} r^{2}}\right)^{2}$, showing an explicit dependence on teacher noise and data corruption. Consequently, higher noise levels require larger amounts of data to retrieve the signal.

\begin{figure}[t!]
    \centering
    \begin{overpic}[width=\linewidth,trim=0 10 0 -10,clip]{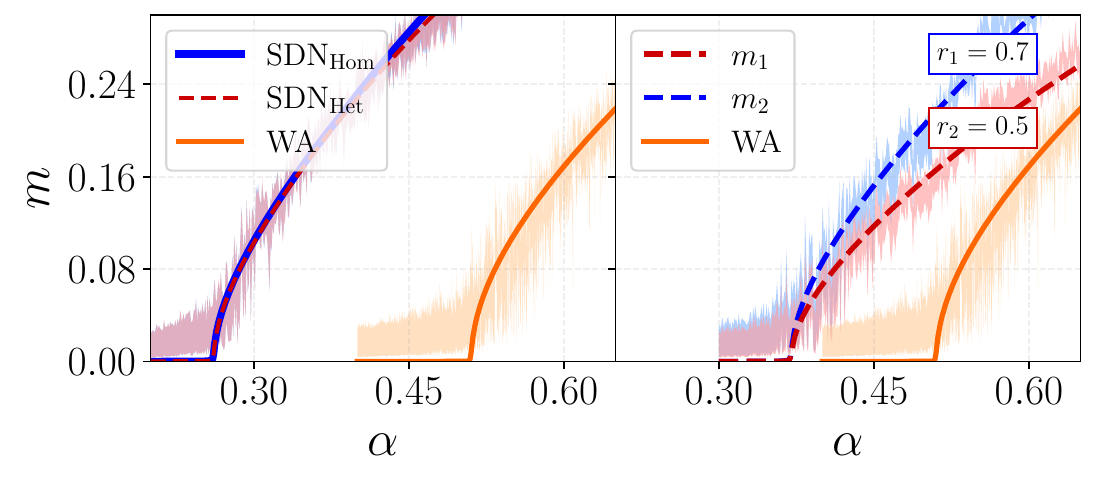}
        \put(15,43){{\figpanel{a}}}
        \put(59,43){{\figpanel{b}}}
    \end{overpic}
    \caption{Comparison of the magnetization obtained from the replica computation (solid and dashed lines) with Monte Carlo (MC) simulations (shaded areas) for a two-student system using the SDN ansatz at $\beta=0.5, \, \hat{\beta}=0.8$. 
    \figpanel{a} SDN-homogeneous ansatz ($r=0.7$) and SDN-heterogeneous ansatz ($r_1=r_2=0.7$) both with $\gamma=1$, compared against the WA baseline at the same corruption rate ($r=0.7$). 
    \figpanel{b} SDN-heterogeneous ansatz with asymmetric corruption ($r_1=0.7, r_2=0.5$) and $\gamma=1$, compared against the WA baseline ($r=0.7$). 
    In both panels, the WA student requires larger values of $\alpha$ to retrieve the teacher. The effect of the coupling in the SDN-homogeneous case significantly reduces the retrieval load. Remarkably, in the SDN-heterogeneous case, the coupling strongly assists both students in finding the teacher, even when one student receives lower-quality samples.}
    \label{fig: simulations}
\end{figure}

\noindent \textbf{(i) SDN.}  To assess the effect of student cooperation we first consider the simplest setting in which all $y$ students experience an identical corruption level: $\bm r \!=\!r \bm 1$. This case is analytically tractable, as all students can be treated symmetrically and characterized by common order parameters $m$ and $q$.
In this homogeneous corruption scenario, as shown in Fig.~\ref{Fig:PD_WA}--(b), increasing the number of collaborating students $y$ systematically enlarges the sR region in the phase diagram: the P-sR transition shifts toward higher temperatures (with the critical temperature increasing with the number of interacting students), while the SG-sR boundary moves toward smaller values of $\alpha$.
The main effect of increasing the number of collaborating students is a continuous shift of the triple point's 
$\alpha_c$ (marked by the pentagons in Fig.~\ref{Fig:PD_WA}-\figpanel{b}), beyond which all students become magnetized simultaneously.
For a single student ($y\!=\!1$), $\alpha_c$ coincides with the WA case in the presence of noise $r$.
As the number of students increases, in the limit $y \to \infty$, an expansion of the system equations around the P-sR boundary (SM) analytically shows that $\alpha_c$  converges to that of the noise‑free WA case ($r\!=\!1$), indicating that collaboration effectively suppresses the noise in the dataset.

A similar result can be seen when considering the more general setting where students receive datasets with heterogeneous corruption levels, i.e., $\bm r\! \neq r \!\bm 1$. The analytical treatment of this scenario becomes significantly more involved, as it requires tracking the individual magnetizations of each student with the teacher, as well as all pairwise overlaps between students, whose number grows rapidly with $y$. We illustrate the case $y=2$ when  $r_1  \neq r_2$ and leave more complex combination of students for further study.
The interaction is beneficial for both students at any chosen temperature, as each one can retrieve the signal with fewer samples than in the WA case, even though the learning task is harder for the less informed student. 
In Fig.~\ref{fig: simulations}, this effect is clear when comparing an extreme heterogeneous case, e.g. $\bm r=(0.7,0.5)$, with both the WA at $r=0.7$ and its homogeneous counterpart, $r_1=r_2=0.7$. As for the homogeneous corruption, the two students start magnetizing simultaneously. Both signals benefit from the interaction, although the higher noise affecting the second model ($r_2\!=\!0.5$) requires each student to observe more data before entering the sR region compared to the homogeneous case, but still much lower than in isolation (WA). Remarkably, even though each student alone would not have enough information to magnetize towards the teacher configuration, their interaction is beneficial for both.
\begin{figure}[t!]
\label{Fig: heterogeneous dataset size}
    \centering
    \begin{overpic}[width=\columnwidth,trim=0 5 0 8,clip]{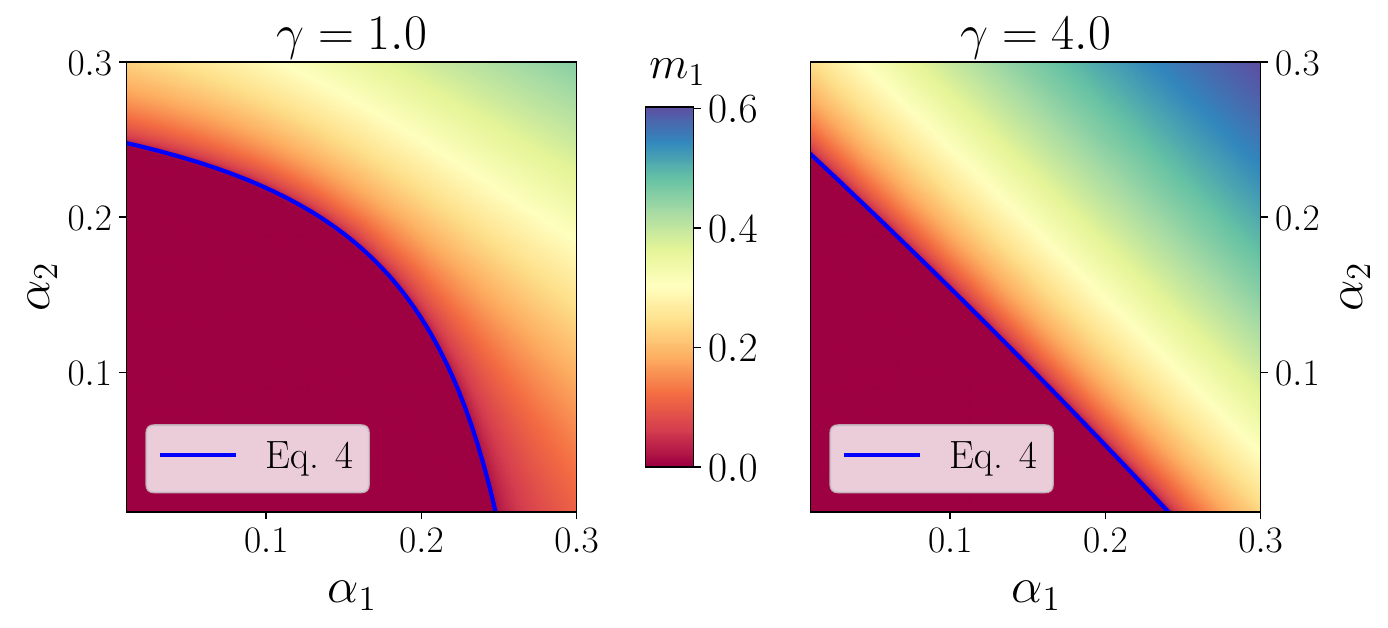}
\put(0,43){{\figpanel{a}}}
\put(53,43){{\figpanel{b}}}
\end{overpic}
    \caption{First student's magnetization ($m_1$) heat-map for the $2$-student interacting system at fixed inverse temperature $\beta=0.5$ with baseline noise $\hat\beta=0.8$ for the SDD scenario. The heterogeneous dataset size effect is described by the magnetization values of the first student in the $(\alpha_1,\alpha_2)$ plane, for two values of $\gamma$. \figpanel{a}: $\gamma=1$; \figpanel{b}: $\gamma=4$. A blue line denotes the sR transition of Eq.\eqref{eq: FL alpha12 transition}. The magnetization of student $u=2$ has the same behavior upon the exchange $\alpha_1 \leftrightarrow \alpha_2$.}\label{Fig: heterogeneous dataset size}
\end{figure}

\noindent \textbf{(ii) SDD} The effect on the sR phase with different dataset sizes is illustrated in Fig.~\ref{Fig: heterogeneous dataset size}. Like in the heterogeneous-SDN we explore the $y \!=\! 2$ case, when $\alpha_1\! \neq \!\alpha_2$. Moreover, to focus on the effects of varying the different $\alpha_u$, we fix the system temperature ($\beta\!=\!0.5$). The plots show the magnetization heatmap of student $1$ around the P-sR transition for $\gamma\!=\!1.0$ (in \figpanel{a}) and $\gamma\!=\!4.0$ (in \figpanel{b}). In both cases, we clearly observe the effect of the coupling between the two students. We can read the WA scenario for student 1 by looking at the $x$-axis. In this case, a non-zero magnetization is reached at $\alpha^{\mathrm{WA}}\!=\! (1-\hat\beta)(1-\beta)/\hat\beta \beta \!\sim\! 0.25$. As shown by the transition line (solid blue curve), as soon as data are introduced for the second student, a non-zero magnetization $m_1>0$ emerges at smaller values of $\alpha_1$.  In this regime, the two students effectively share all the information contained in their datasets, even though the interaction occurs only at the level of the weights and data privacy is preserved. For stronger coupling  ($\gamma=4$) the effect becomes even more pronounced, and the transition line approaches the limiting linear relation $\alpha_1 + \alpha_2 = \alpha^{\mathrm{WA}}$ when $\gamma \to \infty$. In this limiting regime, signal recovery by both students depends only on the total number of samples, rather than how data are distributed between them (i.e., the individual fractions $\alpha_{1,2}$). 
This property extends to an arbitrary number of students in the strong-coupling limit, as shown in Sec.~C.1 of the SM.
For intermediate $\gamma$-values the amount of combined data needed to recover the teacher follows 
\begin{equation}
\label{eq: FL alpha12 transition}
(1-\tanh{(\gamma/2)}^2) k_1 k_2 - k_1 - k_2 + 1 = 0,
\end{equation} 
where $k_u
={\alpha^u \hat\beta \beta}/{[(1-\hat\beta)(1-\beta)]}$ for $u=1,2$.  The transition line is symmetric under the exchange $\alpha_1 \leftrightarrow \alpha_2$, but the magnetization values around the transition are not; they depend on the amount of data each student possesses.

\begin{figure}[t!]
\centering\includegraphics[width=\linewidth]{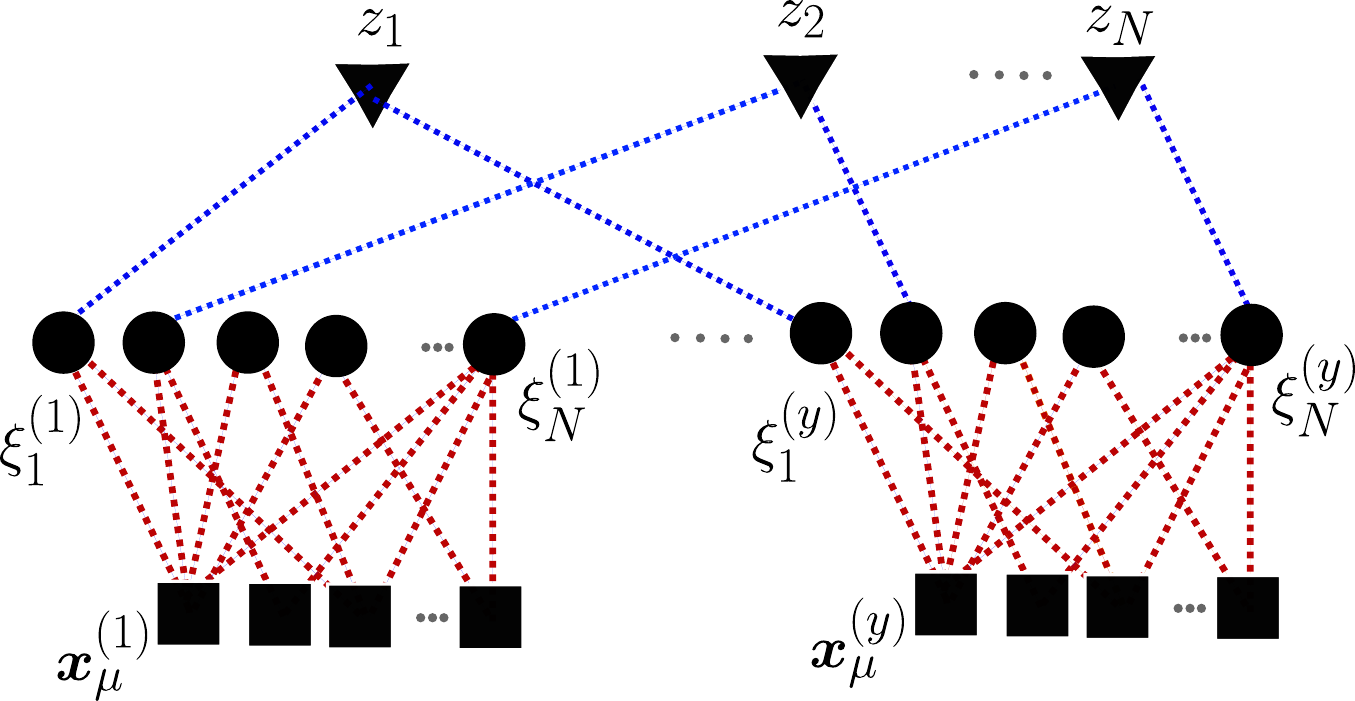}
    \caption{Representation of the RBM associated with the interacting teacher-student model. Each student can be represented as one RBM with weights $\eta_i^{u,\mu}$ (red dotted lines) connecting visible ($\xi_i^u$) and hidden ($x^{(u)}_{\mu}$) units. Each $\xi_i^u$ is further connected via a second hidden unit $z_u$ to the other $u$-RBMs accounting for the cooperation term.}
   \label{fig: structured hidden layer RBM}
\end{figure}

\textbf{Numerical simulations ---} In order to investigate numerically
the shifting effect of the magnetization lines caused by the student interaction, we perform Monte-Carlo (MC) simulations to sample the equilibrium distribution~\eqref{eq:probaBoltz}.
This is achieved by framing the system of interacting Hopfield models in the posterior as an extended RBM,
with a structured hidden layer shown in Fig.~\ref{fig: structured hidden layer RBM}.
In fact, from Eq.(\ref{eq:probaBoltz}), we decouple the two terms inside the Hamiltonian using two Hubbard-Strat\'{o}novich transformations \cite{barra2012equivalence} and obtain a structured bipartite system specified by the following energy function
\begin{equation}
\label{eq: hamiltonian structured RBM}
\mathcal{H} = -\sqrt{\frac{\beta}{N}}\sum_{u,\mu}\left[x_{u\mu}\sum_{i}\xi_{i}^{u}\eta_{i}^{u,\mu}\right]  -\sqrt{\frac{\gamma}{y}}\sum_{i}\left[z_{i}\sum_{u}\xi_{i}^{u}\right]\,.
\end{equation}
The variables $\bm x$ and $\bm z$ are normal distributed and arise, respectively, from the linearization of the term encoding the teacher's data and from the coupled real-replica term in the exponent of Eq.~(\ref{eq:probaBoltz}). The distribution associated with Eq.~\eqref{eq: hamiltonian structured RBM} exhibits a bipartite conditional independence structure: visible units $\bm{\xi}$ are independent given the hidden variables $(\bm{x}, \bm{z})$, and viceversa: this allows to perform simulations through alternate block Gibbs sampling, as commonly used for RBMs \cite{decelle2021equilibrium, mehta2019high}.
In Fig.~\ref{fig: simulations} we show the agreement between MC simulations of a $N\!=\!10^4$  spin system with the theoretical results of the SDN scenario for the 2-interacting students case. The results also enforces the validity of the RS-ansatz to describe the outcome of the many interacting students in the uniform and non-uniform $\bm{r}$ scenarios. The first curves in Fig.~\ref{fig: simulations} (\textbf{a}) are the theoretical predictions for the SDN-homogeneous (solid blue) and SDN-heterogeneous (dashed red), both when $\bm{r}\!=\!0.7\,\bm1$.
Their overlap confirms the consistency of the sR transition description; a result that is further validated by the MC simulations.
We also investigate a pure SDN--heterogeneous case (Fig.~\ref{fig: simulations} (\textbf{b})) with $r_1=0.7 , r_2=0.5$ and confirm further the theoretical results. Additional details about the simulation protocols and results on the SDD scenario are discussed in Sec.~D of the SM. The data that support the findings of this study were generated by numerical simulations. The source code used to generate the simulations is publicly available at \cite{manzan2025code}.

\textbf{Discussion
---} In this work, we have characterized how interactions modify collective inference in federated generative learning. Our results show that cooperation systematically enlarges the retrieval region, reducing the amount of information required for successful reconstruction and improving performance relative to isolated learning.
The mechanisms underlying this improvement depend on the source of heterogeneity. In the student-dependent-noise setting, interactions effectively compensate for noise, progressively recovering the performance of uncorrupted inference as the coupling strength increases, so that students effectively cooperate by averaging out the noise induced onto each one's dataset. In student-dependent dataset, interactions act as a mechanism for data sharing: in the strong-coupling limit, the retrieval threshold depends only on the total sample complexity $\sum_u \alpha_u$, allowing information to be redistributed among students without loss. In both cases, cooperation enables collective recovery in regimes where isolated learners fail, with all students entering the retrieval phase simultaneously.

Beyond these results, the mapping to a RBM with structured hidden interactions provides a physical interpretation of collaborative learning as an equilibrium collective phenomenon, connecting federated inference with energy-based generative models and disordered systems.
Several directions naturally extend this work. Generalizing beyond single-pattern teachers to multi-pattern \cite{theriault2024dense}, sparse, or low-rank structures would establish closer connections with modern machine learning architectures. In such settings, the retrieval phase itself may develop a richer organization, including subdominant metastable states and more complex learning transitions.
Overall, our results suggest that collaboration is not merely a practical constraint imposed by data locality, but can constitute a fundamentally superior inference strategy in heterogeneous, noisy, and data-limited regimes.\\
\begingroup
\renewcommand{\addcontentsline}[3]{} 
\begin{acknowledgments}
The authors acknowledge financial support from grant PID2024-158623NB-C21, funded by MICIU/AEI/10.13039/501100011033 and by ERDF/EU. G.C. acknowledges funding from the Spanish Ministerio de Ciencia, Innovación y Universidades through project María de Maeztu CEX2021-001164-M funded by the MICIU/AEI/10.13039/501100011033. D.T. acknowledges GNFM-Indam. G.M. acknowledge the French ANR grant Scalp (ANR-24-CE23-1320).
\end{acknowledgments}
\endgroup

%

\newpage
\onecolumngrid
\appendix
\appendixtableofcontents
\label{toc}


\section{Derivation of Quenched free energy in homogeneous and heterogeneous systems}
\label{ASec: A}
In this section we discuss how to compute the averaged free energy density for the different federated learning (FL) models discussed in the main text. As we deal with spin-glass models with quenched disorder our solution is obtained by the replica-trick \cite{amit1987statistical,coolen2005theory}, which involve the introduction of a number $n$ of replicas, with the limit $n\to0$ taken afterwards.

In the FL approach we further take into account the presence of ``real'' replicas, corresponding to the interacting students learning the teacher rule. To better identify the difference in replicas we will use the indeces $a,b \in \{1,\cdots,n\}$ to the note the ``fake'' replicas (the ones introduced within the replica-trick framework), while $u,v \in \{1,\cdots,y\}$ are going to index the students.

As a first step we recall the teacher-student setting for this interacting system in a general form that will be further specialized in the student-dependent-noise (SDN) and student-dependent-dataset (SDD) cases.
We first describe the details of the inverse problem of a collection of student Hopfield networks that cooperate and try to recall a teacher rule or signal from another Hopfield machine, which is characterized by one planted pattern $\hat{\bm{\xi}}$ (also referred as Mattis model). The component of the signal are drawned independently from a uniform distribution over $\{-1,1\}$.

The entire system is specified by the state of a collection of 1D vectors
$\bm \xi ^u$, with $u=1,\cdots,y$,  and $\bm\xi^u=(\xi_1^u, \cdots,\xi_N^u)\in \{-1,1\}^N$, representing the inference result of each of the $y$-students.
The probability of each student state follows the posterior
\begin{eqnarray}
    P(\bm{\xi}|\bm{\mathcal{S}}) & ={Z}^{-1}(\bm{\mathcal{S}})\exp\left(\frac{\beta}{N}\sum_{u=1}^{y}\sum_{i<j}^{N}\sum_{\mu_u=1}^{M^u}s_{i}^{\mu_u}s_{j}^{\mu_u}\xi_{i}^{u}\xi_{j}^{u}+\frac{\gamma}{y}\sum_{i}\sum_{u<v}\xi_{i}^{u}\xi_{i}^{v}\right)\,,\label{eq: students posterior1}
    \end{eqnarray}
where a ferromagnetic coupling $\gamma$ induces cooperation between each component of the system.
To identify the signal $\bm{\hat\xi}$, each student have access to its specific dataset $\mathcal{S}^u=\{\bm s^{\mu_u}\}_{\mu_u=1,\cdots,M^u}$ s.t. $M=\sum_{u=1}^{y} M^{u}$ ,which is generated by the teacher itself and represent the set of patterns of each student. The teacher rule cannot be identified directly from each dataset since it is hidden by an overall noise, derived by the intrinsic inverse temperature of the teacher $\hat\beta<1$. Consequently we assume
the student patterns are independent but with a specific spatial correlation induced by the planted
configuration, i.e., they are distributed according to
\begin{eqnarray}
P(\bm{\mathcal{S}}|\bm{\hat\xi})=  \prod_u P(\mathcal{S}^{u}|\bm{\hat\xi})  =  \prod_u \prod_{\mu_u=1}^{M_u}P(\bm{s}^{\mu_u}|\bm{\hat\xi})=
\prod_u\prod_{\mu_u=1}^{M_u}z(\hat\beta)^{-1}\ \exp\left(\frac{\hat{\beta}}{N}\sum_{i<j}^{N} s_{i}^{\mu_u} s_{j}^{\mu_u}\hat{\xi}_{i} \hat{\xi}_{j}\right)
.\nonumber 
\end{eqnarray}

We further induce a student-specific corruption, occurring for each $u$-dataset, which is student-dependent. We structure this via a new set of random variables $\bm{\chi}^{\mu u} \in \{+1,-1\}^N$
\begin{equation}
 \label{Aeq: corruption distribution} 
P(\chi_{i}^{u\mu_u})=\frac{1+r^{u}}{2}\delta(\chi_{i}^{\mu_u}-1)+\frac{1-r^{u}}{2}\delta(\chi_{i}^{\mu_u}+1)\,,
\end{equation}
so that the corrupted example is given by the Hadamard (element-wise) product $\bm\eta^{\mu_u}:=\bm{s}^{\mu_u} \circ \bm{\chi}^{\mu_u}$. This results in a new rephrased distribution for the patterns
\begin{eqnarray}
P(\bm{\mathcal{J}}|\bm{\hat\xi})  =  \prod_u \prod_{\mu_u=1}^{M_u}P(\mathcal{\bm{J}}^{\mu_u}|\bm{\hat\xi})=\prod_u \sum_{\bm \chi_u} \sum_{\{\bm s^{\mu_u}\}}P(\bm\chi^u)P(\{\bm s^{\mu_u}\}|\hat{\bm\xi})\,,
\end{eqnarray}
also the posterior changes, since now the students are trained over the new set $\bm{\mathcal{J}} := \set{\bm{\eta}^{\mu_u}}_{\mu_u=1,\ldots,M^u}^{u=1,\ldots,y}$,
\begin{eqnarray}
P_{}(\bm{\xi}|\bm{\mathcal{J}}) & ={Z}_{c}^{-1}(\bm{\mathcal{J}})\exp\left(\frac{\beta}{N}\sum_{u=1}^{y}\sum_{i<j}^{N}\sum_{\mu_u=1}^{M^u}\eta_{i}^{\mu_u}\eta_{j}^{\mu_u}\xi_{i}^{u}\xi_{j}^{u}+\frac{\gamma}{y}\sum_{i}\sum_{u<v}\xi_{i}^{u}\xi_{i}^{v}\right) \,.
\label{eq: student posterior2}
\end{eqnarray}
Our objective is to compute the disorder-averaged free energy density as a function of the control parameters $\Lambda=\{\beta,\hat{\beta},\bm\alpha,\bm{r}\}$, with $\bm r$ the vector containing the corruption rates $\set{r^u}_{u=1}^y$ and $\bm\alpha = \set{\alpha_u}_{u=1}^y$ being the data load for each student $\alpha^u = M^u/N$, which remains constat in the thermodynamic limit. Denoting with $\left[\cdots\right]^{\bm{\mathcal{J}}}$ the average over the disorder the free energy takes the form
\begin{eqnarray}
-\beta f(\Lambda)=\lim_{N\to\infty}\frac{1}{N}\left[\ln Z\right]^{\bm{\mathcal{J}}}
\label{eq: disordered free energy}
\end{eqnarray}
with partition function
\begin{eqnarray}
Z(\bm{\mathcal{J}}):=\mathbb{E}_{\left\{ \boldsymbol{\xi}^{u}\right\} }\exp\left(\frac{\beta}{N}\sum_{u=1}^{y}\sum_{i<j}^{N}\sum_{\mu_u=1}^{M^u}\eta_{i}^{\mu_u}\eta_{j}^{\mu_u}\xi_{i}^{u}\xi_{j}^{u}+\frac{\gamma}{y}\sum_{i}\sum_{u<v}\xi_{i}^{u}\xi_{i}^{v}\right)\,.\label{eq: quenched partition function}
\end{eqnarray}
It is interesting to note that using the Gauge transformation $\xi_i^u\to\xi_i^u \hat\xi_i $ and $s_i^{\mu_u}\to s_i^{\mu_u} \hat\xi_i$ the set of examples $\bm{\mathcal{S}}$ can be considered as equivalently extracted from a Curie-Weiss model at temperature $\hat\beta$. Using the shorthand notation
\begin{equation}
    Q(\bm\xi^1,\bm\xi^2)=\frac{\bm\xi^1\cdot \bm\xi^2}{N}
\end{equation}
for the overlap between two generic replicas, the average of any bounded function $f(Q(\bm\xi^u , \hat{\bm\xi}))$ becomes
\begin{eqnarray*}
[f(Q(\bm{\xi},\bm{\xip}))]^{\bm{\mathcal{J}}} & = & 2^{-N}\sum_{\bm{\xip},\bm{\mathcal{J}},\bm{\xi}}f(Q(\bm{\xi},\bm{\xip})){P}(\bm{\xi}|\bm{\mathcal{J}},\bm{\xip})P(\bm{\mathcal{J}}|\bm{\hat{\xi}})\\
 & = & \sum_{\bm{\mathcal{J}},\bm{\xi}}f(Q(\bm{\xi},\bm{1})){P}(\bm{\xi}|\bm{\mathcal{J}},\bm{1})P(\bm{\mathcal{J}}|\bm{1})\\
 & = & \sum_{\bm{\mathcal{J}},\bm{\xi}}f(Q(\bm{\xi})){P}(\bm{\xi},\bm{\mathcal{J}})P^{CW}_{\hat\beta}(\bm{\mathcal{J}}) \\
 & = & \meanv{f(Q(\bm{\xi}))}^{\mathcal{\bm{\mathcal{J}}}},
 \end{eqnarray*}
where in the last two passages we drop the conditional dependence on $\bm 1$.
Then the averaged partition function becomes
\begin{eqnarray}
\label{Aeq: general Z^n}
\langle Z^{n}\rangle^{\bm{\mathcal{J}}} =\sum_{\bm{\mathcal{J}}}P_{\hat{\beta}}^{CW}(\bm{\mathcal{J}})\ Z^{n}(\bm{\mathcal{J}})
 & = & \prod_u\sum_{\boldsymbol{\chi^u}}P(\bm\chi^u)\prod_{\mu_u=1}^{M^u}\frac{1}{z(\hat{\beta})}\exp\left(\frac{\hat{\beta}}{2N}\sum_{i,j}s_{i}^{\mu_u}s_{j}^{\mu_u}\right)\ Z^{n}(\bm{\mathcal{J}})\,,\label{}\\ 
Z^{n}(\bm{\mathcal{J}})&=&\prod_{a=1}^n \sum_{\bm \xi^{(a)}}\exp\left(\frac{\beta}{2N}\sum_{u}\sum_{i,j}\eta_{i}^{\mu_u}\eta_{j}^{\mu_u}\xi_{i}^{(a)u}\xi_{j}^{(a)u}\right)\,.
\end{eqnarray}
Eq.(\ref{Aeq: general Z^n}) is general and includes all the FL cases, namely the corruption variation and dataset realizations:
\begin{itemize}
    \item SDN: We consider the teacher generates one realization of the dataset for all students, which is successively corrupted with a rate $r^u$. Thus the indexes $\mu$ and $u$ are independent and
    \begin{equation}
    P^{CW}_{\hat\beta}(\mathcal{\bm J})= \prod_u \Bigg[
\sum_{\boldsymbol{\chi^u}} P(\bm\chi^u)\prod_{\mu=1}^{M}\frac{1}{z(\hat{\beta})}\exp\left(\frac{\hat{\beta}}{2N}\sum_{i,j}s_{i}^{\mu}s_{j}^{\mu}\right) \Bigg].
\end{equation}
\item SDD: When setting $P(\bm\chi^u)=\delta(\bm\chi^u- \bm 1)$, we focus on the case where each student has its own version of the data.
\end{itemize}
\section{Federated Learning in Student Dependent Noise (SDN) scenario}
\label{ASec: federated Learning in Student Dependent Noise (SDN) scenario}
We begin by describing the corruption realization, specified by $\bm \alpha =\alpha \bm 1$ and a generic corruption vector $\bm r$ . In this case the samples of the students are distributed according to
   \[
    P^{CW}_{\hat\beta}(\mathcal{\bm J})= 
\sum_{\set{\boldsymbol{\chi^u}}} P(\set{\boldsymbol{\chi^u}})\sum_{\mathcal{S}}\prod_{\mu=1}^{M}\frac{1}{z(\hat{\beta})}\exp\left(\frac{\hat{\beta}}{2N}\sum_{i,j}s_{i}^{\mu}s_{j}^{\mu}\right) \,.
    \]
After some algebra Eq.(\ref{Aeq: general Z^n}) take the following form
\begin{eqnarray}
\label{}
\langle Z^{n}\rangle^{\bm{\mathcal{J}}} &=&  \sum_{\bm{\mathcal{J}}} P^{CW}_{\hat\beta}(\bm{\mathcal{J}}) Z^n(\bm{\mathcal{J}})
\\
 & = & \sum_{ \set{\bm\xi^{u(a)}}} \ \left(\frac{2^{N}}{z(\hat{\beta})}\overline{e^{\hat{\beta}/2N\sum_{i,j}s_{i}s_{j}+\beta/2N\sum_{a,u}\sum_{i,j}\eta_{i}^u \eta_{j}^u\xi_{i}^{(a)}\xi_{j}^{(a)}}}^{\ \boldsymbol{\eta}^u}\right)^{M}\label{}\,,
\end{eqnarray}
where now the $\overline{\cdots}^{\ \boldsymbol{\eta}^u}=\overline{\cdots}^{\ \bm s \ \boldsymbol{\chi}^u}$ denotes the average of a representative corrupted sample for each student. In particular the one over $\bm s$ involve all the student d.o.f. at once. The exponential under this average can be expanded introducing $n \times y +1$ Gaussian distributions, so that it is transformed to
\begin{equation}
    e^{\hat{\beta}/2N\sum_{i,j}s_{i}s_{j}+\beta/2N\sum_{a,u}\sum_{i,j}\eta_{i}^u \eta_{j}^u\xi_{i}^{(a)}\xi_{j}^{(a)}} = \int Dz \prod_{a,u=1}^{n,y}Dz^{au}\ \overline{e^{\sqrt{\beta/N}\sum_{i}\sum_{a u}z^{u(a)}\xi_{i}^{u(a)}\eta_{i}^u+\sqrt{\hat{\beta}/N}\sum_{i}zs_{i}}}^{\ \boldsymbol{\eta}^u}\,.
\end{equation}
By first averaging over $\bm s$ and then using the expansion $\ln \cosh x \approx \frac{x^2}{2}$,
only the expectation over $P(\bm\chi^u)$ remains
\[
\simeq\overline{\exp\left\{ \sum_{i}\frac{\beta}{2N}\left(\sum_{au}\chi_{i}^{u}z^{au}\xi_{i}^{ua}\right)^{2}+\frac{\hat{\beta}}{2N}\sum_{i}z^{2}+\frac{\sqrt{\beta\hat{\beta}}}{N}z\sum_{i}\sum_{a}\chi_{i}^{u}z^{au}\xi_{i}^{au}\right\} }^{\boldsymbol{\chi^u}}
\]
\[
	\approx\exp\left\{ \frac{\hat\beta}{2}z^2 +\sum_{i}\frac{\beta}{2N}\sum_{\stackrel{a,b}{u\neq v}}r^{u}r^{v}z^{au}\xi_{i}^{ua}z^{bv}\xi_{i}^{vb} +\frac{\sqrt{\beta\hat{\beta}}}{N}z\sum_{iu}r^{u}\sum_{a}z^{au}\xi_{i}^{ua}+\sum_{i}\frac{\beta}{2N}\sum_{\stackrel{a,b}{u\neq v}}r^{u}r^{v}z^{au}\xi_{i}^{ua}z^{bv}\xi_{i}^{vb}\right\} \,.
\]
We introduce the set of order parameters: magnetization of each student with respect to the teacher, $m_u^{a}$, in replica $a$, and the pairwise overlap between the parameters of students $u$ and $v$ in replicas $a$ and $b$,
\begin{equation}
    m^a_u = \frac{1}{N}\sum_{i}\hat\xi_i \xi_i^u \,, \qquad q^{ab}_{uv} = \frac{1}{N} \sum_i \xi_i^{au} \xi_i^{bv},
\end{equation}
and the shorthand notation 
\begin{align}
\det\big(\bm{\Xi}(\bm{q},\bm{m})\big) & :=\int\prod_{au}^{n,y}Dz_{au}Dz\exp\left(\frac{\beta}{2}\sum_{\stackrel{a,b}{u=v}}z^{au}q_{uv}^{ab}z^{bv}+\frac{\hat{\beta}}{2}z^{2}+\sqrt{\beta\hat{\beta}}z\sum_{ua}r^{u}z^{au}m^{ua}+\sum_{i}\frac{\beta}{2N}\sum_{\stackrel{a,b}{u\neq v}}r^{u}r^{v}z^{au}q_{uv}^{ab}z^{bv}\right) \\
&:=\int\prod_{au}^{n}Dz_{au}Dz\exp\Bigg( \frac{\beta}{2}\sum_{\stackrel{a\text{\ensuremath{\neq}}b}{u=v}}z^{au}q_{uv}^{ab}z^{bu}+\frac{\beta}{2}\sum_{a,u}\left(z^{au}\right)^{2}+\frac{\hat{\beta}}{2}z^{2}+\sqrt{\beta\hat{\beta}}z\sum_{ua}r^{u}z^{au}m^{ua}+ \notag\\
& \hspace{8cm}+\frac{\beta}{2}\sum_{\stackrel{a\neq b}{u\neq v}}r^{u}r^{v}z^{au}q_{uv}^{ab}z^{bv}+\frac{\beta}{2}\sum_{\stackrel{a}{u\neq v}}r^{u}r^{v}z^{au}q_{uv}^{aa}z^{av}\Bigg) \,.
\end{align}
Enforcing the definitions of the above order parameters by the appropriate density of states $\mathcal{D}(\bm{m}, \bm q)$ the partition function takes the appropriate variational principle form
\begin{equation}
    \langle Z^n  \rangle^{\bm{\mathcal{J}}} = \int d\bm q d\bm m d\bm{\hat{q}} d\bm{\hat m} e^{-\beta N \hat f(\bm q, \bm m ;\bm{\hat{q}} , \bm{\hat m}) },
\end{equation}
with
\begin{align}
\label{Aeq: corruption general free energy}
    -\beta \hat f(\bm q, \bm m ;\bm{\hat{q}} , \bm{\hat m}) & =  \ln \sum_{\set{\xi^{u(a)}}} \exp\left(\sum_{a,b,u,v\in \mathcal P}\hat{q}_{uv}^{ab}\xi_{a}^{u}\xi_{b}^{v}+\sum_{au}\hat{m}^{au}\xi_{au}+\frac{\gamma}{y}\sum_{a}\sum_{u<v}\xi_{a}^{u}\xi_{a}^{v}\right)+ \notag \\
    &-\sum_{a,b,v,u\in \mathcal P}\hat{q}^{ab}q^{ab}-\sum_{ua}\hat{m}^{au}m^{au} -\frac{\alpha}{2}\ln\det\big(\bm{\Xi}(\bm{q},\bm{m})\big) \,,
\end{align}
where $\mathcal P$ counts the number of independent overlaps  ${ny}\choose{2}$ and we factorize over the number if components $i=1,\cdots, N$. Following the replica trick the actual form for the free energy is given by
\begin{equation}
    -\beta f(\beta, \hat \beta, \alpha)= \lim_{N\to \infty} \frac{1}{Ny} \langle \ln Z \rangle^{\bm{\mathcal J}} = \lim_{\substack{N\to \infty\\ n \to 0}}\frac{\ln\langle Z^n \rangle^{\bm{\mathcal J}}}{N y n}
\end{equation}
\begin{equation}
\label{Aeq: limiting free energy}
 -\beta f \approx \lim_{\substack{N\to \infty\\ n \to 0}}  \frac{1}{N y n} \ln \left( e^{N \Extr \hat f }\right) \approx \lim_{n\to0} \frac{1}{y n} \Extr \hat f(\bm q, \bm m ;\bm{\hat{q}} , \bm{\hat m}) \,.
\end{equation} 
To proceed further, it is natural to choose an ansatz of the solution and all will depend on the corruption we decide to impose to the system. We split the analysis of the corrupted case in two: first we focus on $\bm r = r \bm 1$, where we consider each student is provided with a dataset realized with the same corruption level, then we focus on the 2-student case $\bm r=\set{r_1,r_2} $. 

\subsection{Homogeneous corruption}
Starting from Eq.~(\ref{Aeq: corruption general free energy}), we adjust the set of order parameters: $m_u^{a}$,  and $q^{ab}_{uv}$, to respect a proper ansatz. Since the noise ratio acts in the same way for all the students: $\bm r = r \bm 1$ we provide the following choice for the overlap matrix:
\begin{equation}
\label{Aeq: K nxy overlap matrix}
    \mathcal{K}=\left(\begin{array}{cccc}
Q_{11} & Q_{12} & \cdots & Q_{1n}\\
Q_{21} & Q_{22} & \cdots & Q_{2n}\\
\vdots & \ddots & \cdots\\
Q_{n1} & Q_{n2} & \cdots & Q_{nn}
\end{array}\right)
\end{equation}
with the $y\times y$ $\bm Q$ matrices being: $Q_{aa}^{uv}= \delta_{uv} +t(1-\delta_{uv})$ and $Q_{ab}^{uv}= q\bm 1_{uv} $.
The choice reflects the replica symmetry between the different students which are interacting with realizations of the datasets descending by the same corruption level $r$. 
We expect the average correlation between the students (in the same replica) to be the same due to the fully-
connected topology Hamiltonian in (\ref{eq: student posterior2}). To reflect the two distinct natures of the replicas we introduce the two distinct values $q$ and $t$. Full replica symmetry further allows for a uniform magnetization $m_u^a = m$, $\forall a,u$. This choice simplify the structure of $\det\big(\bm{\Xi}(\bm{q},\bm{m})\big)$ by reducing it to a block form, thereby facilitating the computation. Since we are interested in the $n \to 0$ limit
\begin{align}
    \ln \det\big(\bm{\Xi}(\bm{q},\bm{m})\big) &\approx n\Bigg[\left(y-1\right)\left\{ \ln\left(1-\beta+\beta q-\beta r^{2}\left(q-t\right)\right)+\frac{\beta q\left(r^{2}-1\right)}{1-\beta+\beta q-\beta r^{2}\left(q-t\right)}\right\} +\nonumber \\
    & \hspace{1cm}+\ln\left(1-\beta+\beta q+\left(y-1\right)\beta r^{2}\left(q-t\right)\right)+\frac{\beta q\left(r^{2}-1\right)-y\left(\beta r^{2}q+\frac{\hat{\beta}\beta r^{2}m^{2}}{(1-\hat{\beta})}\right)}{1-\beta+\beta q+\left(y-1\right)\beta r^{2}\left(q-t\right)}\Bigg]\,.
\end{align}
After taking the replica limit, the last expression on the r.h.s. of (\ref{Aeq: limiting free energy}) results in the extremization of the following RS-function
\begin{align}
-\beta f_{RS}(\bm{q},\bm{m};\bm{\hat{q}},\bm{\hat{m}}) &=-\frac{y(y-1)}{2}\hat{t}t+y^{2}\frac{1}{2}\hat{q}q-y\hat{m}m
-\frac{\alpha}{2}\Bigg[\left(y-1\right)\Bigg\{ \ln\left(1-\beta+\beta q-\beta r^{2}\left(q-t\right)\right)+ \notag
\\
& \hspace{0.5cm} +\frac{\beta q\left(r^{2}-1\right)}{1-\beta+\beta q-\beta r^{2}\left(q-t\right)}\Bigg\} +\ln\left(1-\beta+\beta q+\left(y-1\right)\beta r^{2}\left(q-t\right)\right)+
\notag \\
& \hspace{0.5cm}+\frac{\beta q\left(r^{2}-1\right)-y\left(\beta r^{2}q+\frac{\hat{\beta}\beta r^{2}m^{2}}{(1-\hat{\beta})}\right)}{1-\beta+\beta q+\left(y-1\right)\beta r^{2}\left(q-t\right)}\Bigg]
-y\frac{\hat{q}}{2}+ 
\notag \\
& \hspace{0.5 cm}+\left\langle \ln\left[\sum_{\xi^{u}}\exp\left\{\sum_{u}\xi^{u}\left(\hat{m}+z\sqrt{\hat{q}}\right)+\frac{\text{1}}{2}\left(\widehat{t}-\hat{q}+\frac{\gamma}{y}\right)\sum_{u\neq v}\xi^{u}\xi^{v}\right\} \right]\right\rangle _{z}\,,
\end{align}
where $\meanv{.}_{z}$ denotes the expectation w.r.t. a standard Gaussian
random variable $z\sim\mathcal{N}(0,1)$.
The solution of the variational principle is reached by the following values of the order parameters
\begin{align}
\hat{m}&=\alpha\frac{\hat{\beta}\beta r^{2}m}{\Delta_{T}\Delta_{y}},\label{Aeq:self-consistency} \\
\hat{q} & =\frac{\alpha \beta^2}{\Delta^2 y^2}(y-1) q (1-r^2)^2 + \frac{\alpha \beta^2}{y^2 \Delta_y^2} \left(1+(y-1)r^2 \right) \left( y \frac{\hat \beta r^2 m^2}{\Delta_T} +q(1+r^2(y-1))\right),\\
\hat{t}&=\frac{\alpha}{y}\left[ \frac{\Delta-\Delta_y}{\Delta \Delta_y} + \frac{\beta r^2}{\Delta_y^2}y(\frac{\beta \hat\beta r^2 m^2}{\Delta_T}+\beta r^2 q) \right]+ \frac{\alpha}{y} \beta^2 q r^2 (1-r^2)\frac{(\Delta-\Delta_y)(\Delta+\Delta_y)}{\Delta^2 \Delta_y^2}, 
\end{align}

\begin{align}
    m&=\left\langle\frac{1}{y}\mathbb{E}_{\xi^u|z} \left[\sum_u \xi^u\right]\right\rangle_z\,, \\
q&=\left\langle\frac{1}{y^2}\mathbb{E}^2_{\xi^u|z} \left[\sum_u \xi^u\right]\right\rangle_z\,, \\
t & =\left\langle\frac{1}{y(y-1)}\mathbb{E}_{\xi^u|z} \left[ \sum_{u<v} \xi^u \xi^v\right] \right\rangle_z,
\end{align}
with the following definitions
\begin{align}
\Delta_T &= 1-\hat\beta ,\\
\Delta &= 1-\beta+\beta q +\beta r^2 (t-q) ,\\
\Delta_y &= 1-\beta +\beta q-\beta r^2 (y-1)(t-q) \,.
\end{align}
The average $\mathbb{E}_{\xi^u|z} \left[\, \cdot \,\right]$ stands for the expectation over
\begin{equation}
P(\{\xi^u\}) \exp{\left\{ \sum_u \xi^u \left( \hat m +z\sqrt{\hat q} \right) + \sum_{u<v} \left(\frac{\gamma}{y} + \hat t -\hat q \right) \xi^u \xi^v \right\} },
\label{Aeq: Homog-CW}
\end{equation}
which is the Boltzmann measure over a Curie-Weiss model of $y$ spins with uniform coupling $J = \frac{\gamma}{y} + \hat t -\hat q$ and external field $ h(z) = \hat m +z\sqrt{\hat q}  $. $P(\{\xi^u\})$ stands for the product measure of the binary distribution for all the students’ patterns. 

\subsubsection{$y \to \infty$ limit}
As illustrated in Fig.(\ref{Fig:PD_WA}) the more students are interacting the better is the signal retrieval effect of the system. In particular the coupled system is able to clean the superimposed corruption, going towards the critical load of the clean dataset. The aim of this section is to analyze the $y \to \infty$ limit to show this property.
We begin by
examining the averaging distribution (\ref{Aeq: Homog-CW})
\begin{equation}
P(\{\xi^u\}) \exp{\left\{ \sum_u \xi^u \left( \hat m +z\sqrt{\hat q} \right) + \sum_{u<v} \left(\frac{\gamma}{y} + \hat t -\hat q \right) \xi^u \xi^v \right\} }\,.
\end{equation}
The first terms inside the exponential and the one proportional to $\gamma/y$ are extensive in $y$, while the others are of $\mathcal{O}(y^2)$. To preserve the linear scaling w.r.t. the system size we assume the combination $\hat t - \hat q = \hat\delta /y$. The same assumption applies to
the respective conjugate combination. As a consequence we investigate the extremization (\ref{Aeq: limiting free energy}) when $y\to \infty$. First we identify the relevant $\mathcal{O}(y)$ terms in $\hat f(\bf q, \bf m;\bf{\hat q},\bf{\hat m})$, then impose again the previous RS ansatz. By using the following expansions
\begin{itemize}
\centering
    \item $\ln\left(\Delta\right)=\ln\left(1-\beta+\beta q+\beta r^{2}\frac{\delta}{y}\right)\approx \ln\left(1-\beta+\beta q\right)$
    \item $\Delta^{-k} \approx\left(1-\beta+\beta q\right)^{-k}$
    \item $\ln\left(\Delta_{y}\right)=\ln\left(1-\beta+\beta q-\left(y-1\right)\beta r^{2}\frac{\delta}{y}\right)\approx \ln\left(1-\beta+\beta q-\beta r^{2}\delta\right)$
    \item $\Delta_{y}^{-k}\approx \left(1-\beta+\beta q-\beta r^{2}\delta\right)^{-k}$\,,
\end{itemize}
we arrive at the ($y$-intensive) form of the free energy for the $y \to \infty$ case:
\begin{align}
\label{Aeq: Inf free energy}
-\frac{1}{y}\beta f_{RS}(\bm{q},\bm{m};\bm{\hat{q}},\bm{\hat{m}}) & =-\frac{\hat{\delta}q}{2}-\frac{\hat{q}\delta}{2}+\frac{\hat{q}q}{2}-\hat{m}m-\frac{\alpha}{2}\Bigg\{\ln\left(1-\beta+\beta q\right)+\frac{\beta q\left(r^{2}-1\right)}{1-\beta+\beta q}\Bigg\}+
\notag \\ & \hspace{0.5cm}+\frac{\alpha}{2}\Bigg[\frac{\left(\beta r^{2}q+\frac{\hat{\beta}\beta r^{2}m^{2}}{\Delta_T}\right)}{1-\beta+\beta q-\beta r^{2}\delta}\Bigg]-\frac{\hat{q}}{2}+\notag\\
  & \hspace{0.5cm}  -\frac{J \bar M^2}{2} + \log 2+\log \cosh\left( h(z) + J \bar M \right) \,.
\end{align}
For the final line of (\ref{Aeq: Inf free energy}) we use the Curie-Weiss result in the thermodynamic limit with system size $y$. Then
\begin{equation}
\label{Aeq: YInf distribution}
\sum_{\xi^{u}}\exp\left\{ \sum_{u}\xi^{u}\left(\hat{m}+z\sqrt{\hat{q}}\right)+\frac{\text{1}}{2y}\left(\hat{\delta}+\gamma\right)\sum_{u\neq v}\xi^{u}\xi^{v}\right\} = \exp\left\{ y\ \textbf{Extr}_M \left[ -\frac{JM^2}{2} + \log 2+\log \cosh\left( h(z) + J M \right) \right] \right\} \,,
\end{equation}
where we identify $J=\hat{\delta}+\gamma$ and $h(z)=\hat m + z \sqrt{\hat q}$. The extremum is reached for
\begin{equation}
\bar M = \tanh \left( h(z) + J \bar M \right)\,.
\end{equation}
With the above identification we can easily rewrite the set of equations
for the order parameters (\ref{Aeq:self-consistency})

\begin{equation}
\label{Aeq: YInf magnetization}
m=\left\langle\frac{1}{y}\mathbb{E}_{\xi^u|z} \left[\sum_u \xi^u\right]\right\rangle_z = \int Dz \bar M\,,
\end{equation}

\begin{equation}
q=\left\langle\frac{1}{y^2}\mathbb{E}^2_{\xi^u|z} \left[\sum_u \xi^u\right]\right\rangle_z = \int Dz \bar M^2\,,
\end{equation}

\begin{equation}
\label{Aeq: YInf delta}
\delta=\left\langle\frac{1}{y(y-1)}\mathbb{E}_{\xi^u|z} \left[ \sum_{u<v} \xi^u \xi^v\right] -\frac{1}{y^2}\mathbb{E}^2_{\xi^u|z} \left[\sum_u \xi^u\right]\right\rangle_z=\int Dz \frac{\left( 1- \bar M^2 \right)^2 J}{1-J\left(1- \bar M^2 \right)}\,
\end{equation}

\begin{align}
    \hat{m}&=\alpha\frac{\hat{\beta}\beta r^{2}m}{\Delta_{T}\left( 1-\beta+\beta q-\beta r^{2}\delta\right)}\\
\hat{q}&=\frac{\alpha\beta r^{2}}{\left( 1-\beta+\beta q-\beta r^{2}\delta\right)^{2}}\left(\beta r^{2}q+\hat{M}\right) \\
\hat{\delta}&=\hat{q}-\alpha\left[\frac{\beta r^{2}}{ 1-\beta+\beta q}+\frac{\beta^{2}q\left(1-r^{2}\right)}{\left( 1-\beta+\beta q\right)^{2}}\right]+\alpha\left[\frac{\beta r^{2}}{ 1-\beta+\beta q-\beta r^{2}\delta}-\frac{\beta\left(\beta r^{2}q+\hat{M}\right)}{\left( 1-\beta+\beta q-\beta r^{2}\delta\right)^{2}}\right]\,,
\end{align}
where we used the shorthand notation $\hat M =\frac{\hat{\beta}\beta r^{2}m^{2}}{\Delta_T} $. 
Now we can proceed with a small order parameter expansion around zero. This is because at very high temperature $(\beta \sim 0)$ the distribution (\ref{Aeq: YInf distribution}) has no external effective fields and therefore both magnetization and q-overlap are in a paramagnetic phase. Lowering the temperature may start a spin glass or signal retrieval transition, with nonzero $m$ and $q$. Assuming the transition is continuous we linearize(\ref{Aeq: YInf magnetization}-\ref{Aeq: YInf delta}), together with their conjugate counterparts. Since every order parameter depends on $\bar M$, it is sufficient to obtain its expression for small values $\bar{M}\approx\frac{\hat{m}+z\sqrt{\hat{q}}}{\left(1-J\right)}$ and substitute inside all the other equations. This leads to the following lines:
\begin{itemize}
\centering
    \item[]\textbf{P-sR} line: \begin{equation}
        1=\frac{1}{\left(1-J\right)}\alpha\frac{\hat{\beta}\beta r^{2}}{\Delta_{T}\left(1-\beta-\beta r^{2}\delta\right)}\,,\end{equation}
    \item[]\textbf{P-SG} line: \begin{equation}
        1=\frac{\alpha\beta^{2}r^{4}}{\left(1-J\right)^{2}\left(1-\beta-\beta r^{2}\delta\right)^{2}}\,.\end{equation}
\end{itemize}
 The intersection between the two  lines is the triple point containing the critical load $(\alpha_c,T_c)$, where $\alpha_{c}=\left({\Delta_{T}}/{\hat{\beta}}\right)^{2}$, which is exactly the same of $y$-Hopfield interacting models without corruption. For $\bm r= \bm 1$ this threshold is independent by the number of interacting students, $y\geq 1$ or $y\to\infty$ \cite{amsdottorato12369}.

\subsection{Heterogeneous corruption}
In this section, we specialize the previous replica computation to a more intricate setting in which two students ($y=2$) are trained on two versions of the dataset with different corruption rates, $r_{1} \neq r_{2}$. As a consequence, the two students are no longer equivalent and replica symmetry between them must be broken. In this case, the overlap matrix is given by
\begin{equation}
\label{Aeq: K nxy overlap matrix}
    \mathcal{K}=\left(\begin{array}{cccc}
Q_{11} & Q_{12} & \cdots & Q_{1n}\\
Q_{21} & Q_{22} & \cdots & Q_{2n}\\
\vdots & \ddots & \cdots\\
Q_{n1} & Q_{n2} & \cdots & Q_{nn}
\end{array}\right)
\end{equation}
with the following ansatz for each of the $2\times2$ internal matrices
\begin{equation}
    Q_{aa}=\left(\begin{array}{cc}
1 & t\\
t & 1
\end{array} \right)\,, \qquad Q_{ab}=\left(\begin{array}{cc}
q_{1} & q\\
q & q_{2}
\end{array}\right)\,.
\end{equation}
 Within this assumption we can rewrite the former generic free energy equation (\ref{Aeq: corruption general free energy}) as
\begin{align}
\label{eq: free energy corrupted 2 students}
    \beta\hat{f}(\bm{q},\bm{m};\bm{\hat{q}},\bm{\hat{m}})&=	\ln\sum_{\xi^{1},\ldots,\xi^{n}}\exp\left(\hat{t}\sum_{a}\;\xi_{a}^{1}\xi_{a}^{2}+\hat{q}\sum_{a\neq b}\xi_{a}^{1}\xi_{b}^{2}+\frac{\hat{q_{1}}}{2}\sum_{a\neq b}\xi_{a}^{1}\xi_{b}^{1}+\frac{\hat{q_{2}}}{2}\sum_{a\neq b}\xi_{a}^{2}\xi_{b}^{2}\right.+\nonumber\\
    &+
	\left.\hat{m}_{1}\sum_{a}\xi_{a}^{1}++\hat{m}_{2}\sum_{a}\xi_{a}^{2}+\frac{\gamma}{2}\sum_{a}\xi_{a}^{1}\xi_{a}^{2}\right)-\frac{\alpha}{2}\ln\det\big(\bm{\Xi}(\bm{q},\bm{m})\big) + \\\nonumber
&-	n\hat{t}t+\frac{1}{2}n\hat{q_{1}}q_{1}+\frac{1}{2}n\hat{q_{2}}q_{2}+n\hat{q}q-n\hat{m_{1}}m_{1}-n\hat{m_{2}}m_{2} \nonumber
\end{align}
with the same number of independent parameters $\binom{ny}{y}\Big|_{y=2}=2n^{2}-n$. 
We linearize the exponential term over the replica index  by adding and subtracting the following quantities $\pm\frac{\hat{q}_{1}}{2}\sum_{a}\left(\xi_{a}^{1}\right)^{2}\pm\frac{\hat{q}_{1}}{2}\sum_{a}\left(\xi_{a}^{2}\right)^{2}\pm\hat{q}\sum_{a}\xi_{a}^{1}\xi_{a}^{2}$. This allows for a more general Hubbard-Stratonovich transformation \cite{theriault2024dense,centonze2024statistical}  via complex fields $\omega$, $\bar \omega$
\begin{align}
    \notag
        \sum_{\left\{ \boldsymbol{\xi}^{au}\right\} }e^{-\frac{\hat{q}_{1}}{2}n-\frac{\hat{q}_{2}}{2}n}&\exp\left(\hat{t}\sum_{a}\;\xi_{a}^{1}\xi_{a}^{2}+\frac{\hat{q}_{1}}{2}\left(\sum_{a}\xi_{a}^{1}\right)^{2}+\frac{\hat{q}_{2}}{2}\left(\sum_{a}\xi_{a}^{2}\right)^{2}+\hat{q}\sum_{a}\xi_{a}^{1}\sum_{b}\xi_{b}^{2}-\hat{q}\sum_{a}\xi_{a}^{1}\xi_{a}^{2}+\right.\\&+\hat{m}_{1}\sum_{a}\xi_{a}^{1}++\hat{m}_{2}\sum_{a}\xi_{a}^{2}+\frac{\gamma}{y}\sum_{a}\xi_{a}^{1}\xi_{a}^{2}\left.\right),\\
        \label{eq: entropic term linear replica}
        \sum_{\left\{ \boldsymbol{\xi}^{au}\right\} }e^{-\frac{\hat{q}_{1}}{2}n-\frac{\hat{q}_{2}}{2}n}&\exp\left(\left.\hat{t}\sum_{a}\;\xi_{a}^{1}\xi_{a}^{2}+\hat{m}_{1}\sum_{a}\xi_{a}^{1}++\hat{m}_{2}\sum_{a}\xi_{a}^{2}+\frac{\gamma}{y}\sum_{a}\xi_{a}^{1}\xi_{a}^{2}-\hat{q}\sum_{a}\xi_{a}^{1}\xi_{a}^{2}\right)\right.\times \notag\\ 
        \times&\int Dz_{1}Dz_{2}\exp\left(z_{1}\sqrt{\hat{q}_{1}}\sum_{a}\xi_{a}^{1}+z_{2}\sqrt{\hat{q}_{2}}\sum_{a}\xi_{a}^{2}\right)\int \frac{d\omega\,d\bar{\omega}}{\pi}\exp\left(-\omega\bar{\omega}+\sqrt{\hat{q}}\omega\sum_{a}\xi_{a}^{1}+\sqrt{\hat{q}}\bar{\omega}\sum_{a}\xi_{a}^{2}\right).
    \end{align}

Now lets focus on the computation of $\det\big(\bm{\Xi}(\bm{q},\bm{m})\big)^{-1/2}$: 
    \begin{align}
        \det\big(\bm{\Xi}(\bm{q},\bm{m})\big)^{-1/2}&=\int \prod_{\substack{a,b\\ u,v}\in\mathcal{P}} \mathcal{D}z^{au}\Delta_T^{-1/2}\exp \Big[ \frac{\beta}{2}\Big( \sum_{\substack{a,b\\u}}z^{au}z^{bu} q^{ab}_{uu}+\sum_{\substack{a,b\\u\neq v}} r^u r^v q^{ab}_{uv}\Big)+\frac{\hat\beta \beta}{2 \Delta_T}\sum_{\substack{a,b\\u,v}}r^u r^v m^{a}_u m^b_v \ z^{au}z^{bv}\Big] 
    \end{align}

\begin{equation}
\label{Aeq: SDN-heter overlap matrix}
\boldsymbol{\Xi}(\boldsymbol{q},\boldsymbol{m})=
\left(
\begin{array}{cccc}
\begin{pmatrix}
M_1 & T\\
T   & M_2
\end{pmatrix}
&
\begin{pmatrix}
Q_1 & Q\\
Q   & Q_2
\end{pmatrix}
& \cdots &
\begin{pmatrix}
Q_1 & Q\\
Q   & Q_2
\end{pmatrix}
\\[1.5em]

\begin{pmatrix}
Q_1 & Q\\
Q   & Q_2
\end{pmatrix}
&
\ddots
&
&
\vdots
\\[1.5em]

\vdots
&
&
\ddots
&
\begin{pmatrix}
Q_1 & Q\\
Q   & Q_2
\end{pmatrix}
\\[1.5em]

\begin{pmatrix}
Q_1 & Q\\
Q   & Q_2
\end{pmatrix}
&
\cdots
&
\begin{pmatrix}
Q_1 & Q\\
Q   & Q_2
\end{pmatrix}
&
\begin{pmatrix}
M_1 & T\\
T   & M_2
\end{pmatrix}
\end{array}
\right).
\end{equation}

with the following definitions for the terms inside the blocks
\begin{align}
    M_1=1-\beta-\frac{\hat\beta \beta}{\Delta_T}(r_1 m_1)^2 &\quad M_2=1-\beta-\frac{\hat\beta \beta}{\Delta_T}(r_2 m_2)^2, \\ 
     Q_1=-\beta q_1-\frac{\hat\beta \beta}{\Delta_T}(r_1 m_1)^2 &\quad Q_2=-\beta q_2-\frac{\hat\beta \beta}{\Delta_T}(r_2 m_2)^2,
    \\
    T=-r_1 r_2 \beta \ t &-\frac{\hat\beta \beta}{\Delta_T}(r_1 m_1)(r_2 m_2), \\
    Q=-r_1 r_2 \beta \ q &-\frac{\hat\beta \beta}{\Delta_T}(r_1 m_1)(r_2 m_2).
\end{align}
The above matrix structure allows us to obtain the following expression for $\log \det \left( \bm \Xi \right)$:
\begin{align}
\label{eq: logDet}
    \frac{1}{n}\log \det \left( \bm \Xi \right) &= \log \det{^{(1)}}\left( \bm \Xi \right)+\mathcal{O}(n)\\& =\frac{M_2 Q_1 + (M_1 - 2 Q_1) Q_2 - 
 2 Q (T - Q)}{(M_1 - Q_1) (M_2 - Q_2) - (Q - T)^2} + \log\left[ (M_1 - Q_1) (M_2 - Q_2) - (Q - T)^2 \right]
      +\mathcal{O}(n)\,,
\end{align}
where the label $^{(1)}$ represent the leading order in $n$. 

Finally the approximated form of the free energy (\ref{eq: free energy corrupted 2 students}) is
\begin{align}
\label{eq: final free energy corrupted 2 students}
    -\beta{f}_{RSB}(\bm{q},\bm{m};\bm{\hat{q}},\bm{\hat{m}})&=\Bigg\langle 
    \ln \sum_{\xi^{1},\xi^{2}}\exp\left( \xi^1 \left( \hat m_1 +z_1 \sqrt{\hat q_1}+\omega \sqrt{\hat q} \right)+\xi^2 \left( \hat m_2 +z_2 \sqrt{\hat q_2}+\bar\omega \sqrt{\hat q} \right) \right.+\\
    &+
	\left. \xi^1 \xi^2 \left( \hat t - \hat q +\frac{\gamma}{2} \right) \right) \Bigg\rangle_{G's}-\frac{\alpha}{2}\log \det{^{(1)}}\left( \bm \Xi \right) + \\
&-	\hat{t}t+\frac{1}{2}\hat{q_{1}}q_{1}+\frac{1}{2}\hat{q_{2}}q_{2}+\hat{q}q-\hat{m_{1}}m_{1}-\hat{m_{2}}m_{2}-\frac{1}{2}\hat{q_{1}} -\frac{1}{2}\hat{q_{2}} \,, 
\end{align}   
where we denote with $\langle \cdot \rangle_{G's}$ the integration over all the Gaussian variables in Eq.(\ref{eq: entropic term linear replica}) and we substitute the $\mathcal{O}(1)$ expansion of Eq.(\ref{eq: logDet}).

\subsubsection{Order parameters' equations}
By renaming the quantities $h_1(z_1,\omega) = \hat m_1 +z_1 \sqrt{\hat q_1}+\omega \sqrt{\hat q} $, $h_2(z_2,\bar\omega) =\hat m_2 +z_2 \sqrt{\hat q_2}+\bar\omega \sqrt{\hat q}$ and $J=\hat t - \hat q +\frac{\gamma}{2} $ it is possible to obtain the new  version of Eqs.(\ref{Aeq:self-consistency}) 

\begin{align}
\label{Aeq: heter-SDN m1}
    m_1 &= \big\langle \mathbb{E}_{\{\xi^{1,2}\}|G's} \left[ \xi^1 \right]\big\rangle_{G's}\,,\\
    m_2 &= \big\langle \mathbb{E}_{\{\xi^{1,2}\}|G's} \left[ \xi^2 \right]\big\rangle_{G's}\,,\\
    q_1 &= \big\langle \Big(\mathbb{E}_{\{\xi^{1,2}\}|G's}[ \xi_1 ] \Big)^2 \big\rangle_{G's}\,,\\
    q_2 &= \big\langle \Big(\mathbb{E}_{\{\xi^{1,2}\}|G's}  [ \xi_2 ]\Big)^2 \big\rangle_{G's}\,,
    \\
    t &= \big\langle \mathbb{E}_{\{\xi^{1,2}\}|G's} [ \xi_1 \xi_2 ]\big\rangle_{G's}\,,\\
    q &= \big\langle \mathbb{E}_{\{\xi^{1,2}\}|G's} [\xi_1]\ \mathbb{E}_{\{\xi^{1,2}\}|G's}[\xi_2 ] \big\rangle_{G's}\,. \label{Aeq: heter-SDN q}
\end{align}
with the averaging distribution (\ref{Aeq: Homog-CW}) changed to
\begin{align}
    \label{eq: corr-distribution}
    P(\{\xi^1,\xi^2\})\exp\left\{ h_1 (z_1,\omega) \ \xi^1  +h_2 (z_2,\omega) \ \xi^2+ J\ \xi^1 \xi^2 \right\}\,,
\end{align}
and the outer average over the set of Gaussian variables $G's$.\\

\subsubsection{Reality of the order parameters and reduction to two Gaussian integrals}

The order parameters as written involve a four-dimensional Gaussian integral 
over $(z_1, z_2, u, v)$ and a complex-valued integrand due to the presence of 
$\omega = u+iv$ and $\bar\omega = u-iv$. We show that the imaginary parts vanish 
and that the four integrals reduce to two.

We take the the magnetization $m_1$ as reference example. We can rewrite it as in the following:
\begin{equation}
m_1 = \left\langle \frac{\sum_{\xi_1,\xi_2} \xi_1 R_{\xi_1\xi_2}
[\cos\phi + i\sin\phi]}{\sum_{\xi_1,\xi_2} R_{\xi_1\xi_2}
[\cos\phi + i\sin\phi]} \right\rangle,
\end{equation}
where $R_{\xi_1\xi_2} := \exp(a_1\xi_1 + a_2\xi_2 + J\xi_1\xi_2)$ and 
$\phi := v\sqrt{\hat{q}}(\xi_1 - \xi_2)$. Separating real and imaginary parts,
\begin{equation}
\mathcal{R}(m_1) = \left\langle \frac{
\left(\sum \xi_1 R \cos\phi\right)\left(\sum R \cos\phi\right)
+\left(\sum \xi_1 R \sin\phi\right)\left(\sum R \sin\phi\right)
}{
\left(\sum R \cos\phi\right)^2 + \left(\sum R \sin\phi\right)^2
} \right\rangle,
\end{equation}
\begin{equation}
\mathcal{I}(m_1) = i\left\langle \frac{
\left(\sum \xi_1 R \sin\phi\right)\left(\sum R \cos\phi\right)
-\left(\sum \xi_1 R \cos\phi\right)\left(\sum R \sin\phi\right)
}{
\left(\sum R \cos\phi\right)^2 + \left(\sum R \sin\phi\right)^2
} \right\rangle.
\end{equation}
Since $\sin\phi$ is odd in $v$ while $\cos\phi$ is even, the imaginary part 
$\mathcal{I}(m_1)$ is an odd function of $v$ and therefore vanishes under the 
Gaussian average. The same parity argument applies to $m_2$, $q_1$, $q_2$, 
$t$, and $q$, so all order parameters are real.

To reduce the four integrals to two, we perform the linear change of variables 
$(u, v, z_1, z_2) \to (S, D, A_1, A_2)$ defined by
\begin{equation}
u = \frac{S+D}{2\sqrt{\hat{q}}}, \qquad
v = \frac{i(D-S)}{2\sqrt{\hat{q}}}, \qquad
z_1 = \frac{A_1 - S}{\sqrt{\hat{q}_1}}, \qquad
z_2 = \frac{A_2 - D}{\sqrt{\hat{q}_2}}.
\end{equation}
Under this transformation the inner sum 
$\sum_{\xi_1,\xi_2}\exp(\xi_1(\hat{m}_1+A_1)+\xi_2(\hat{m}_2+A_2)+J\xi_1\xi_2)$
becomes independent of $S$ and $D$, so these variables can be integrated out. 
The result is a two-dimensional Gaussian integral over $(A_1, A_2)$ with 
covariance matrix
\begin{equation}
\Sigma = \begin{pmatrix} \hat{q}_1 & \hat{q} \\ \hat{q} & \hat{q}_2 \end{pmatrix},
\end{equation}
yielding for instance
\begin{equation}
m_1 = \int \frac{dA_1\,dA_2}{2\pi\sqrt{\hat{q}_1\hat{q}_2 - \hat{q}^2}} 
\exp\!\left[-\frac{A_1^2\hat{q}_2 - 2A_1 A_2\hat{q} + A_2^2\hat{q}_1}
{2(\hat{q}_1\hat{q}_2-\hat{q}^2)}\right] 
\frac{\sum_{\xi_1,\xi_2} e^{\xi_1(\hat{m}_1+A_1)+\xi_2(\hat{m}_2+A_2)
+J\xi_1\xi_2}\,\xi_1}
{\sum_{\xi_1,\xi_2} e^{\xi_1(\hat{m}_1+A_1)+\xi_2(\hat{m}_2+A_2)+J\xi_1\xi_2}},
\end{equation}
and analogously for all other order parameters, with the replacements
$\langle\cdot\rangle_G \to \langle\cdot\rangle_A$ and 
$h_u(z_u,\omega)\to \hat{m}_u + A_a$.

For numerical evaluation, the correlated Gaussian over $(A_1,A_2)$ is sampled 
via the following decomposition
\begin{equation}
A_1 = \sqrt{\hat{q}_1}\,g_1, \qquad 
A_2 = \sqrt{\hat{q}_2}\!\left(\rho\, g_1 + \sqrt{1-\rho^2}\,g_2\right), \qquad
\rho = \frac{\hat{q}}{\sqrt{\hat{q}_1\hat{q}_2}},
\end{equation}
with $g_1,g_2$ independent standard Gaussians. This yields the six 
estimators
\begin{align}
m_1 &= \langle \tilde{\mathbb{E}}[\xi_1] \rangle_{g_1,g_2}, 
& m_2 &= \langle \tilde{\mathbb{E}}[\xi_2] \rangle_{g_1,g_2}, \\
q_1 &= \langle (\tilde{\mathbb{E}}[\xi_1])^2 \rangle_{g_1,g_2}, 
& q_2 &= \langle (\tilde{\mathbb{E}}[\xi_2])^2 \rangle_{g_1,g_2}, \\
t   &= \langle \tilde{\mathbb{E}}[\xi_1\xi_2] \rangle_{g_1,g_2}, 
& q   &= \langle \tilde{\mathbb{E}}[\xi_1]\,\tilde{\mathbb{E}}[\xi_2] \rangle_{g_1,g_2},
\end{align}
where each conditional expectation $\tilde{\mathbb{E}}[\ \!\cdot\!\ ]$ is taken over the Boltzmann weight 
$\exp(\xi_1 h_1 + \xi_2 h_2 + J\xi_1\xi_2)$ with $h_u = \hat{m}_u + A_u$ 
and $J = \hat{t} - \hat{q} + \gamma/2$.


\subsubsection{Conjugate order parameters }
The conjugate order parameters come from the the $\ln\det^{(1)}\!\big(\bm{\Xi}\big)$ contribution inside the free energy (\ref{eq: final free energy corrupted 2 students}). We can consider this term as the generating functional of the conjugated variables. To show this let reshape it as in the following
\begin{equation}
\label{Aeq: conjugate params generating functional}
\mathcal{F}
\equiv
\frac{A(x)}{B(x)} + \log B(x),
\qquad
x \in \{m_u,q_u,q,t\}.
\end{equation}
We introduce the shorthand
\begin{equation}
\mathcal{M}_u \equiv M_u - Q_u = 1-\beta(1-q_u),
\qquad
\mathcal{Q} \equiv Q-T = -r_1 r_2 \beta(q-t),
\end{equation}
whose original definitions were given in
(\ref{Aeq: SDN-heter overlap matrix}). The numerator and denominator appearing in the $\ln\det^{(1)}\!\big(\bm{\Xi}\big)$--generating (\ref{Aeq: conjugate params generating functional})   are
\begin{equation}
A
=
M_2Q_1 + M_1Q_2
- 2Q_1Q_2
- 2Q(T-Q),
\qquad
B
=
\mathcal{M}_1 \mathcal{M}_2 - \mathcal{Q}^2
=
\frac{\Delta_{\mathrm{dn}}}{\Delta_T},
\end{equation}
where
\begin{equation}
\Delta_{\mathrm{dn}}
:=
\Delta_T
\Big[
1 + \beta(-2 + q_1 + q_2)
+ \beta^2
\big(
1 + q_1(q_2-1) - q_2
- r_1^2 r_2^2 (q-t)^2
\big)
\Big].
\end{equation}
The combination $B-A$ reads
\begin{equation}
B-A
=
M_1 M_2 - T^2
- 2M_1Q_2
- 2M_2Q_1
+ 3Q_1Q_2
- 3Q^2
+ 4QT.
\end{equation}
A generic conjugate variable is obtained from the saddle-point condition
\begin{equation}
\hat{x}
=
\gamma_x\,\partial_x \mathcal{F},
\end{equation}
with
\begin{equation}
\partial_x \mathcal{F}
=
\frac{
(\partial_x A)\,B
+
(\partial_x B)(B-A)
}{
B^2
},
\label{Aeq:master_derivative formula}
\end{equation}
and prefactors
\begin{equation}
\gamma_x =
\begin{cases}
-\alpha/2, & x \in \{m_1,m_2,t\}, \\[0.3em]
\alpha, & x \in \{q_1,q_2\}, \\[0.3em]
\alpha/2, & x = q.
\end{cases}
\end{equation}
The partial derivatives of $B$ are
\begin{equation}
\partial_{q_1} B = \beta\mathcal{M}_2, \qquad
\partial_{q_2} B = \beta\mathcal{M}_1, \qquad
\partial_t B = \partial_q B = -2\beta r_1 r_2\,\mathcal{Q},
\end{equation}
and of $A$:
\begin{equation}
\partial_{q_1} A = \beta(2Q_2 - M_2), \qquad
\partial_{q_2} A = \beta(2Q_1 - M_1), \qquad
\partial_t A = 2\beta r_1 r_2\, Q, \qquad
\partial_q A = 2\beta r_1 r_2\, T.
\end{equation}
The conjugate magnetizations and overlaps take the compact form
\begin{align}
\label{Aeq: heter-SDN hat m1}
\hat{m}_1 &= \frac{\alpha\beta\beta_T r_1^2
\left[\mathcal{M}_2\,m_1 - \mathcal{Q}\,r_2^2\,m_2\right]}{\Delta_{\mathrm{dn}}}, \\[0.8em]
\hat{m}_2 &= \frac{\alpha\beta\beta_T r_2^2
\left[\mathcal{M}_1\,m_2 - \mathcal{Q}\,r_1^2\,m_1\right]}{\Delta_{\mathrm{dn}}}.\\[0.8em]
\hat{q}_1 &= \frac{\alpha\beta}{B^2}
\left[\mathcal{M}_2(B-A) + (2Q_2-M_2)\,B\right], \\[0.8em]
\hat{q}_2 &= \frac{\alpha\beta}{B^2}
\left[\mathcal{M}_1(B-A) + (2Q_1-M_1)\,B\right],\\[0.8em]
\hat{q} &= \frac{\alpha\beta r_1 r_2}{B^2}
\left[T\cdot B + (Q-T)(B-A)\right], \\[0.8em]
\hat{t} &= \frac{-\alpha\beta r_1 r_2}{B^2}
\left[Q\cdot B + (Q-T)(B-A)\right].
\label{Aeq: heter-SDN hat t}
\end{align}

\subsubsection{Phase diagram for Heterogeneous-SDN }
\begin{figure}[H]
    \centering
    
     
    \begin{minipage}{0.4\linewidth}
        \centering
        \includegraphics[width=\linewidth]{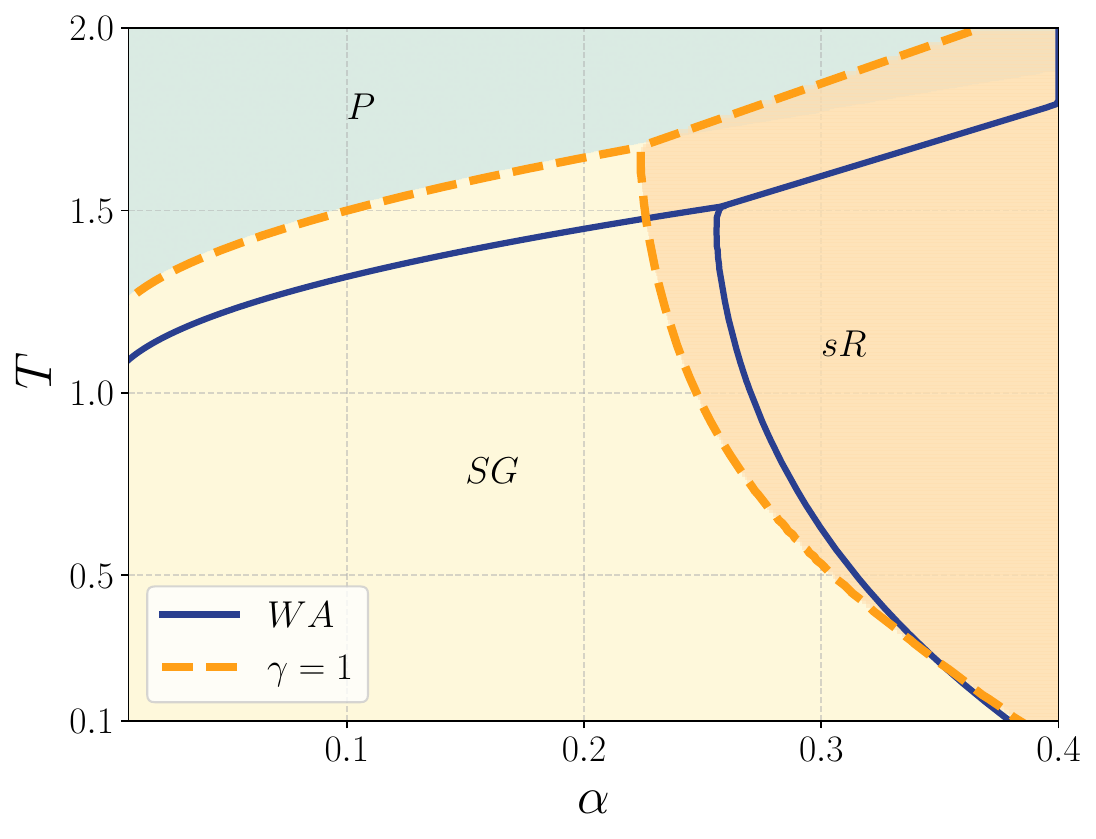}
    \end{minipage}%
    \hspace{0.05\linewidth} 
    \begin{minipage}{0.4\linewidth}
        \centering
        \includegraphics[width=\linewidth]{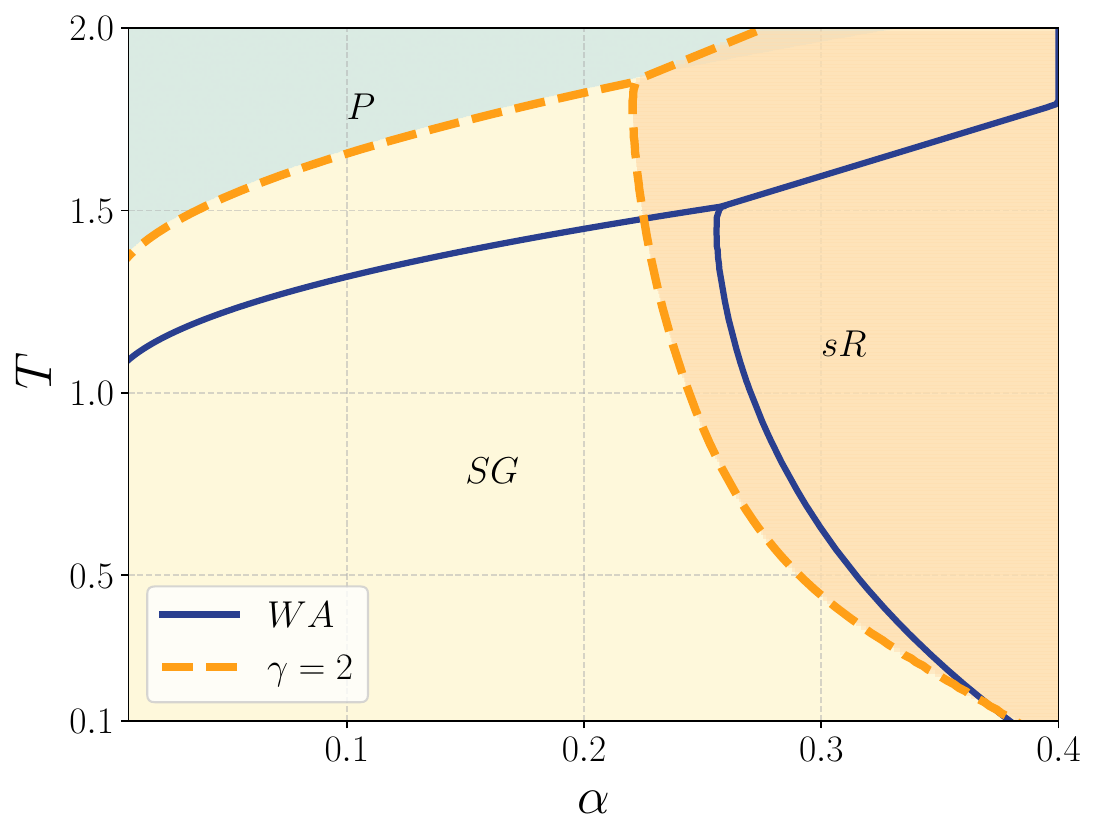}
    \end{minipage}

      \vspace{0.5cm}

    \begin{minipage}{0.4\linewidth}
        \centering
        \includegraphics[width=\linewidth]{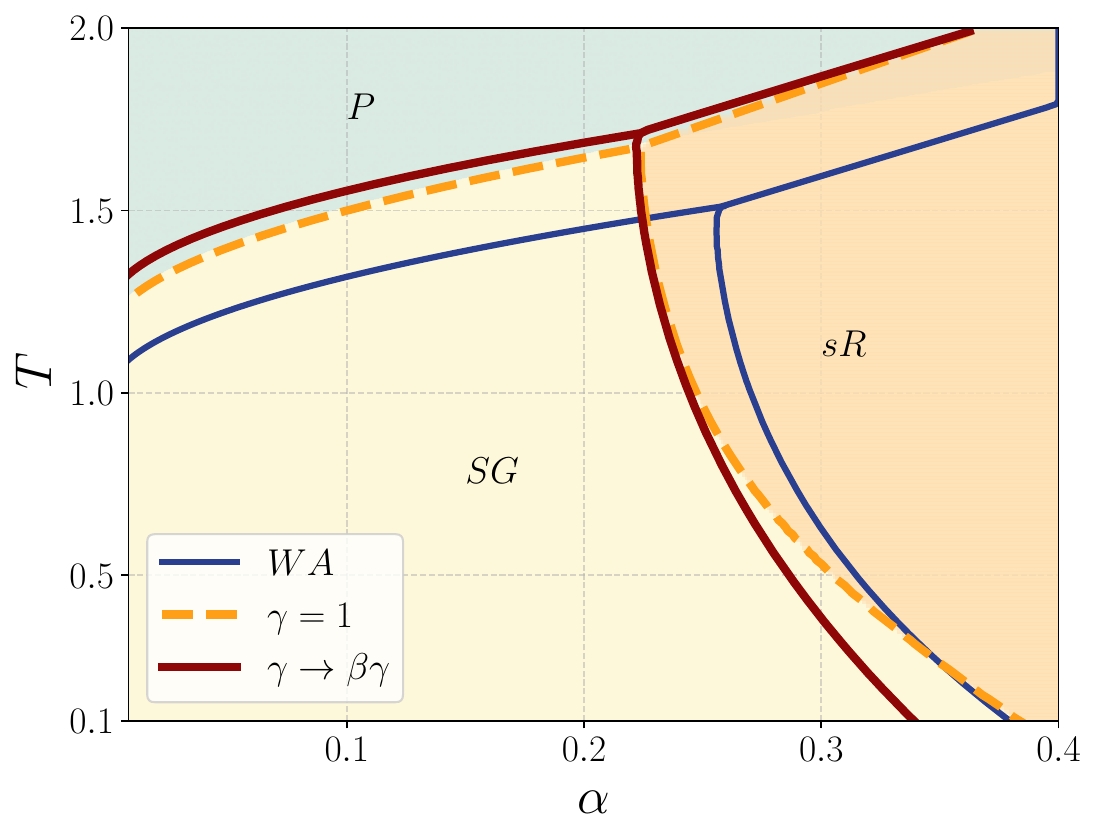}
    \end{minipage}

    \vspace{0.4cm}

    \caption{Phase diagram of the 2-student interacting system with corruption coefficients $r_1=0.7$, $r_2=0.5$ for baseline noise set to $\hat\beta=0.8$. \textbf{Left}: $\gamma =1$. In the paramagnetic (P) and spin-glass (SG) regions, recovering the teacher is impossible. For larger dataset sizes ($\alpha$), the students enter the signal retrieval region (sR), where they learn by generalization. The compharison with the WA case shows the coupled model reduce the critical load $\alpha_c$ w.r.t. the solo $r=0.7$. \textbf{Right}: $\gamma =2$. In this case the sR phase is enlarged, both at high temperature (the transition appears sooner than the $\gamma=1$) and at low temperature. \textbf{Bottom:} On top of the same phase diagram of $\gamma=1$ the red solid curve displays the effect of the rescaling $\gamma \to \beta \gamma$. The stable solution at $T\sim 0$ is not the uncoupled one, as in the previous two images. In this case the data and coupling Hamiltonians have the same scaling and the coupled solution is stable at any temperature.}
    \label{Afig: corrupted_interacting_students}
\end{figure}

The set of equations (\ref{Aeq: heter-SDN m1}-\ref{Aeq: heter-SDN q}) and (\ref{Aeq: heter-SDN hat m1}-\ref{Aeq: heter-SDN hat t}) describe the stable configuration of the free energy (\ref{eq: final free energy corrupted 2 students}). 
In Fig.~\ref{Afig: corrupted_interacting_students} we describe how the students--teacher alignment changes with the aid of phase diagrams. The panels describe the effect of interaction between the two students compared with the WA case for heterogeneous-SDN with $r_1=0.7$, $r_2=0.5$. Each ($\alpha,T$)--phase diagram shows the equilibrium properties of the system, identified by the three phases presented in the main text: Paramagnetic (P) region: $\bm m = 0$; Signal retrieval (sR) region: $\bm m > \bm 0, \bm q > \bm 0$; Spin glass (SG) region: $\bm m = \bm 0, \bm q > \bm 0$.
Only in the sR region  the system is able to learn by generalization; in the other regions, learning is impossible. Each diagram features a triple point $P_c = (\alpha_c,T_c)$ at the intersection of the three phases. The value $\alpha_c$ represents the minimal amount of data required for the system to enter the retrieval regime. In particular, when the amount of noisy data ($\hat\beta<1$) reaches a certain threshold, the students are able to learn by condensing the information contained in the dataset.

 In the WA case, this point has the simple analytical form $\alpha_c^u = \left(\frac{1-\hat{\beta}}{\hat{\beta}r_u^{2}}\right)^{2}$, explicitly depending on the amount of noise (teacher plus corruption) for each student. As expected higher noise requires each student singularly to have more data to enter the sR region. Fig.~\ref{Afig: corrupted_interacting_students} shows the single student ($r=0.7$) as a solid blue line with its triple point at $\alpha_c \sim 0.26$. The one at $r=0.5$ would have $\alpha_c \sim 1$.
The dashed lines refers to the case with non-zero coupling (left: $\gamma=1$, right: $\gamma=2$). The most notable effect is the new position of $\alpha_c$. We clearly assist at the benefit for the second student which reduces notably the amount of data it needs to retrieve the teacher. The first student’s instead has a minor effect, but still evident.  Hence, the net effect of coupling Hopfield students in the heterogeneous-noise scenario is the reduction of the critical amount of data used to retrieve the teacher.
The lower panels of FIG.~\ref{Afig: corrupted_interacting_students} illustrate the thermodynamic consequence of explicitly rescaling the mutual coupling with the inverse temperature, namely $\gamma \to \beta \gamma$, within the same ansatz. In the unscaled regime (top panels), the total exponent inside the partition function Eq.~\eqref{eq: quenched partition function} reads $-\beta \mathcal{H}_{\mathrm{data}}-\gamma\mathcal{H}_{\mathrm{int}}$. Because the quenched data landscape scales with $\beta$ while the interaction remains finite, the local fields completely overwhelm the coupling as $\beta \to \infty$, forcing the students to decouple and recover their independent WA solutions at $T=0$. Conversely, when the interaction is rescaled as $\beta \gamma$, the effective Boltzmann weight balances. This structural shift prevents the zero-temperature fracture; both students undergo a simultaneous phase transition to the retrieval state across all temperatures.

\subsubsection{Transition to retrival states: P--sR}
When a sufficient amount of data is provided, the system of two coupled students may undergo a continuous transition from the paramagnetic phase to a retrieval state upon lowering the temperature. The transition line can be obtained by expanding the saddle--point equations for small magnetizations $m_1,m_2$, while keeping $q=q_\tau \sim 0$.
In this regime, the magnetization equations reduce to
\begin{align}
\label{eq:m1_PsR}
m_1 &= \hat m_1 + \hat m_2\, t ,\\
\label{eq:m2_PsR}
m_2 &= \hat m_2 + \hat m_1\, t ,
\end{align}
where
\begin{align*}
    t &= \Omega_2(\xi^1 \xi^2),
\end{align*}
\begin{align}
\Omega_2(g(\xi^1,\xi^2)) &=
Z_{CW_2}^{-1}
\sum_{\xi^1,\xi^2}
\exp\!\left[\left(\frac{\gamma}{2}+\hat t\right)\xi^1\xi^2\right]
g(\xi^1,\xi^2) .
\end{align}

We introduce the quantities
\begin{align}
\Delta_T &= 1-\hat\beta ,\\
\Delta &= 1-\beta+\beta r_1 r_2 t ,\\
\Delta_y &= 1-\beta-\beta r_1 r_2 t ,
\end{align}
together with
\begin{align}
\delta_1(m_1,m_2) &= m_1(1-\beta)+\beta m_2 r_2^2 t ,\\
\delta_2(m_1,m_2) &= m_2(1-\beta)+\beta m_1 r_1^2 t .
\end{align}
The conjugate parameters read
\begin{align}
\hat m_1 &=
\frac{\alpha \beta \hat\beta r_1^2}{\Delta_T \Delta \Delta_y}
\delta_1(m_1,m_2)
= a_1 m_1 + b m_2 ,\\
\hat m_2 &=
\frac{\alpha \beta \hat\beta r_2^2}{\Delta_T \Delta \Delta_y}
\delta_2(m_1,m_2)
= a_2 m_2 + b m_1 ,\\
\hat t &=
\frac{\alpha \beta \hat\beta (r_1 r_2)^2 t}{\Delta \Delta_y} ,
\end{align}
where
\begin{align}
a_u &=
\frac{\alpha \beta \hat\beta r_u^2 (1-\beta)}
{\Delta_T \Delta \Delta_y},
\qquad
b =
\frac{\alpha \beta^2 \hat\beta (r_1 r_2)^2 t}
{\Delta_T \Delta \Delta_y}.
\end{align}

Substituting these expressions into Eqs.~\eqref{eq:m1_PsR}--\eqref{eq:m2_PsR}, we obtain the linear system
\begin{equation}
M
\begin{pmatrix}
m_1\\
m_2
\end{pmatrix}
=0,
\end{equation}

\begin{equation}
M =
\begin{pmatrix}
1-a_1-bt & -(b+a_2 t)\\
-(b+a_1 t) & 1-a_2-bt
\end{pmatrix}.
\end{equation}
The continuous P--sR transition is signaled by the vanishing of the smaller eigenvalue of $M$. This happen when

\begin{align}
\label{Aeq: dtM=0 p-sR corrupted}
    \det M &= (b^2 -a_1 a_2)t^2 -2 b t + 1-(a_1+a_2)+a_1 a_2 - b^2 =0 \\
    &=(1-a_1-bt)(1-a_2-bt)-(b+a_1 t)(b+a_2 t)=0\,.
    \notag
\end{align}

The relevant eigenvalue of $M$ is related to the magnetization branch continuously connected to the correlated mode. Particularly it signal the transition from the paramagnetic phase towards the direction of both positive magnetized students. 
From a direct calculation
\begin{equation*}
    \lambda_{\pm}=\frac{TrM}{2}\pm \frac{\sqrt{TrM^2-4 \det M}}{2}\,,
\end{equation*}
with the corresponding eigenvector at the transition 
\begin{equation*}
    e_{-}=
\begin{pmatrix}
1\\
\frac{1-a_1-bt}{b+a_2 t} 
\end{pmatrix}.
\end{equation*}
From Eq.(\ref{Aeq: dtM=0 p-sR corrupted}) we see that the ratio in $e_-$ is positive.
This means as soon as one of the student magnetizes, also the other feels a jump of its own magnetization.   

In the absence of student-specific corruption, $r_1=r_2=1$, this eigenvalue reproduces the transition line of the full replica--symmetric case  ($m_u^a=m$ $\forall a,u$, $q_{uv}^{ab}=q$ for $a\neq b$) \cite{amsdottorato12369}
\begin{equation}
\beta
=
\frac{\Delta_y\big|_{r_1=r_2=1}}
{\alpha\,\hat\beta\,\Delta_T\,(1+t)} .
\end{equation}

\subsubsection{Transition to Spin-Glass states: P-SG}
The transition to the Spin-Glass state is analyzed by the stability of the $m_1=m_2=0$ and $q_1,q_2,q \sim0$ solution. In this regime, the pattern components $\xi_i^u$ fluctuate without global bias but undergo a \textit{freezing} transition where they become correlated across replicas.
Under these hypothesis the equations for the overlaps are
\begin{align}
\label{eq: q1 P-sG transition}
    q_1&=\hat{q}_1+\hat{q}_2t^2+2\hat q t\\
    \label{eq: q2 P-sG transition}
    q_2&=\hat{q}_2+\hat{q}_1t^2+2\hat q t\\
\label{eq: q P-sG transition}
q&=\hat{q}(1+t^2)+(\hat{q}_1+\hat{q}_2)t\,,
\end{align}
where the equations for $t,\hat t$ are the same of the previous transition. The remaining conjugated variables equations read as
\begin{align}
\label{eq: C-q1 P-sG transition}
    \hat q_1&=B{q}_1+A{q}_2+2 C q \\
    \label{eq: C-q2 P-sG transition}
    \hat q_2&=B{q}_2+A{q}_1+2 C q \\
\label{eq: C-q P-sG transition}
\hat q&=D{q}+C({q}_1+{q}_2)\,,
\end{align}
whith $A,B,C,D$ all functions of $\Lambda = (\alpha,\beta,r_1,r_2)$ and $t$:
\begin{align*}
    A= \frac{\alpha \beta^2}{(\Delta_y \Delta)^2} (r_1r_2)^2 & t^2 \beta^2 \ ; \quad B = \frac{\alpha \beta^2}{(\Delta_y \Delta)^2} (1-\beta)^2 \ ; \quad C=\frac{\alpha \beta^2}{(\Delta_y \Delta)^2} \beta (1-\beta) (r_1r_2)^2 t  \ ; \\ D&= \frac{\alpha \beta^2}{(\Delta_y \Delta)^2} (r_1r_2)^2 \left[ (1-\beta)^2 +\beta^2(r_1r_2)^2t^2 \right] \ .
 \end{align*}
 As for the P--sR transition, we merge the previous equations in a linear system. Combining Eqs. (\ref{eq: q1 P-sG transition}) and (\ref{eq: C-q1 P-sG transition}) we get

 \begin{equation}
Q \boldsymbol q = \boldsymbol 0 \,,
\end{equation}
\begin{equation}
\label{eq: linear system P-SG}
Q= \begin{bmatrix}
a & b & c\\
b & a & c\\
d & d & e
\end{bmatrix}
\end{equation}
with $a, b, c, d, e$ functions of the corresponding uppercase quantities. One can see that this matrix commutes with the permutation $P$ of the first two rows, i.e. $[P,Q]=0$.
By choosing a suitable basis, $Q$ becomes block-diagonal:
\begin{equation}
Q'= \begin{bmatrix}
a-b & 0 & 0\\
0 & a+b & \sqrt{2}c\\
0 & \sqrt{2}d & e
\end{bmatrix} = \begin{bmatrix}
Q'_{anti} & 0 & 0\\
0 & & \\
0 &  & Q'_{sym}
\end{bmatrix}\,.
\end{equation}
The diagonalization corresponds to a transformation into the basis of orthonormal freezing modes

\begin{equation*}
\label{Aeq: anti_mode SG-corrpt}
 \bm x_{anti}= \frac{1}{\sqrt{2}}(q_1 - q_2)  \quad \bm{x}_{sym} = 
 \begin{pmatrix}
     \frac{1}{\sqrt{2}}(q_1 + q_2) \\ q  
 \end{pmatrix}
 \end{equation*}
The transformation reveals two distinct physical pathways, which emerge from the request of a non-trivial solution of (\ref{eq: linear system P-SG}) i.e. $\det Q'=(a-b)[(a+b)e-2cd]=0$. 
\begin{itemize}
    \item \textbf{The Antisymmetric Branch} ($\det Q'_{anti}=0$): the two students freeze into configurations that are anti-aligned across replicas $(q_1=-q_2)$, while the cross-overlap $q$ remains zero;
    \item \textbf{The Symmetric Branch} ($\det Q'_{sym}=0$): Here, the cross-overlap $q$ is proportional to the sum of the self-overlaps. This mode is driven by the internal pattern correlation $t$. Because $t$ represents the structural similarity of the weights degrees of freedom, any freezing in $q_1$ and $q_2$ must be accompanied by a non-zero $q$ to maintain the energetic favorability of the $\gamma$ term in the Hamiltonian in (\ref{eq: quenched partition function}).
\end{itemize}
 Then the transition is  given by the condition $\det Q'_{\text{sym}}=0$, which yields:
\begin{equation}
\det Q'_{\text{sym}}=(1-t_+(A+B)-4C)(1-4tC-t_+D)-(4t_+C+4tD)(t(A+B)+t_+C)=0 \,,
\end{equation}
where $t_+=1+t^2$. To obtain directly  the physical transition it is possible to combine the equation of the linear system to land directly in the symmetric sector by the following combination 
\begin{equation*}
    q_1+q_2+2q = (\hat q_1+ \hat q_2 +2 \hat q)(1+t)^2
\end{equation*}
where
\begin{align*}
    \phantom{\frac{}{}}\hat q_1+ \hat q_2 +2 \hat q &= (A+B+2C)(q_1+q_2)+ 2q(2C+D) \,; \\
    A+B+2C&=\frac{\alpha \beta^2 }{(\Delta_y \Delta)^2}\Big( \beta^2(r_1 r_2)^2 t^2 +(1-\beta)^2+2\beta(r_1 r_2)^2(1-\beta)t \Big) \;;\\
    2C+D &= \frac{\alpha \beta^2 (r_1 r_2)^2}{(\Delta_y \Delta)^2} \Big( (1-\beta)^2 + \beta^2 (r_1 r_2)^2 t^2 +2(1-\beta)\beta t \Big) \,.
\end{align*}
In the specific no-noise  \cite{amsdottorato12369} $A+B+2C = 2C+D = \Delta\big|_{r1=r2=1}$, giving 
\begin{align*}
    q_1+q_2+2q &= (q_1+q_2+2q) \frac{\alpha \beta^2}{\Delta_y\big|_{r1=r2=1}^2}(1+t)^2\,, \\
    \Rightarrow \beta^2 &=\frac{(1-\beta-\beta t)^2}{\alpha (1+t)^2}\,.
\end{align*}

\section{Federated Learning in Student Dependent Dataset (SDD) scenario}
\label{ASec: federated Learning in Student Dependent Dataset (SDD) scenario}

In this section we perform the evaluation of the second FL scenario described in the main text when the teacher generate a different dataset for each student. When this happen, Eq.(\ref{Aeq: general Z^n}) is still valid. As already mentioned in Sec.\ref{ASec: A} we need to tailor the distribution of the data 
\begin{equation}
\label{Aeq: general distribution 2}
P^{CW}_{\hat\beta}(\mathcal{\bm J}) =\prod_u\sum_{\boldsymbol{\chi^u}}P(\bm\chi^u)\prod_{\mu_u=1}^{M^u}\frac{1}{z(\hat{\beta})}\exp\left(\frac{\hat{\beta}}{2N}\sum_{i,j}s_{i}^{\mu_u}s_{j}^{\mu_u}\right)
\end{equation}
accordingly. To do so we assume that the teacher generates a total of
\[
M=\sum_{u=1}^{y} M^{u}
\]
configurations at temperature $\hat{\beta}$. These samples are then split into $y$ subsets of arbitrary size $M^{u}$, each assigned to student $u$. From the teacher’s perspective, this procedure is equivalent to producing a single dataset of size $M$; however, for the learning dynamics, the subdivision matters because each student conditions its posterior only on its own data.
Following this construction the data distribution from the teacher pattern is
\begin{equation}
\label{Aeq: different dataset disorder distribution}
P(\boldsymbol{S}\mid \hat{\boldsymbol{\xi}})
   = \prod_{u=1}^{y}\prod_{\mu_{u}=1}^{M_{u}}
      \frac{\exp\!\left(
      \frac{\hat{\beta}}{N}\sum_{i<j}
      s_{i}^{\mu_{u}} s_{j}^{\mu_{u}}\, \hat{\xi}_{i}\hat{\xi}_{j}
      \right)}
      {Z_{u}(\hat{\beta})}\, ,
\end{equation}
which corresponds to (\ref{Aeq: general distribution 2}) after the choice $P(\bm\chi^u)=\delta(\bm\chi^u- \bm 1)$.
Given this splitting, the posterior over students’ weights is proportional to
\begin{equation}
P(\boldsymbol{\xi}\mid \boldsymbol{S}) \propto 
\exp\left[
   \frac{\beta}{N}\sum_{u=1}^{y}\sum_{\mu_{u}=1}^{M_{u}}
   \sum_{i<j}s_{i}^{\mu_{u}}s_{j}^{\mu_{u}}\xi_{i}^{u}\xi_{j}^{u}
   + \frac{\gamma}{y}\sum_{i}\sum_{u<v}\xi_{i}^{u}\xi_{i}^{v}
\right].
\end{equation}
The structural change -- other than $\bm r =1$ -- with respect to the first FL case described (SDN) is that the effective load of each student is now
\begin{equation}
\alpha^{u}=\frac{M^{u}}{N}.
\end{equation}
Consequently, within the replica computation the disorder average factorizes over students, and the saddle-point equations 
acquire an explicit dependence on $\alpha^{u}$.
By following the same passages leading to (\ref{Aeq: corruption general free energy}), but now using the distribution (\ref{Aeq: different dataset disorder distribution}) to compute the disordered average of the replica trick, we obtain the $n$-power averaged partition function for this new setup
\begin{align}
\label{Aeq: Mu [Z^n]}
\left[Z^{n}\right]^{\bm{\mathcal{S}}} =\sum_{\{\bm \xi^u\}^a}& \prod_u \left(\frac{2^N}{z(\hat{\beta})} \overline{e^{\hat{\beta}/2N \sum_{i,j}s_i^u s_j^u + \beta/2N\sum_a\sum_{i,j}s_i^u s_j^u \xi_i^{au}\xi_j^{au}}}^{\ \bm{s}^u}\right)^{M^u} \,. \end{align}
Since data are independent for each student and across students, the result of the expectation  over each $\bm s^u$ has the same form of the single Hopfield case \cite{alemanno2023hopfield}:
\begin{equation}
    \frac{2^N}{z(\hat{\beta})} \overline{e^{\hat{\beta}/2N \sum_{i,j}s_i^u s_j^u + \beta/2N\sum_a\sum_{i,j}s_i^u s_j^u \xi_i^{au}\xi_j^{au}}}^{\ \bm{s}^u}  =: \det\big(\bm{\Xi}(\bm{q}^u,\bm{m}^u)\big)^{-1/2}\,,
\end{equation}
where we introduced the order parameters
\begin{equation}
    q_{ab}^u(\bm{\xi}^u)=\frac{1}{N}\sum_{i=1}^N \xi^{au}_i\xi^{bu}_i \,,\ \ \ \ \ m_a^u(\bm{\xi}^u)=\frac{1}{N}\sum_{i=1}^N \xi^{au}_i \;.
\end{equation}
In this case the matrix $\mathcal{K}$ appearing in (\ref{Aeq: K nxy overlap matrix}) has a simple form since only the self overlap $q^{u}_{a,b}$ for $a\neq b$ appears as order parameter. Because of this we deal only with the diagonal entries of $\mathcal{Q}_{ab}=\text{diag}(q^1_{ab},q^2_{ab},\ldots,q^y_{ab})$, which number of independent overlap is $y\frac{n(n-1)}{2}$.
By fixing the order parameters with the use of delta functions, $\left[Z^{n}\right]^{\bm{\mathcal{S}}}$ can be written as an extremization problem, taking $N\to\infty$:
\begin{equation}
    \left[Z^{n}\right]^{\bm{\mathcal{S}}} \approx e^{N\Extr f(\bm{q}^u,\bm{m}^u,\bm{p}^u,\bm{m_0}^u;\bm{\hat{q}}^u,\bm{\hat{m}}^u,\bm{\hat{p}}^u,\bm{\hat{m}_0}^u)}
\end{equation}
with
{\small
\begin{align}
-\beta f(\bm{q}^u,\bm{m}^u;\bm{\hat{q}}^u,\bm{\hat{m}}^u) & =  \ln \sum_{\{\xi^u\}^a} \exp \left( \sum_{a<b}\hat{q}^{ab}_u\xi_{au} \xi_{bu} +\sum_{au} \hat{m}^{au} \xi_{au}+\frac{\gamma}{y}\sum_a \sum_{u<v}\xi_{au}\xi_{bv}\right) +\nonumber \\
    & -\sum_u \frac{\alpha_u}{2} \ln \det\big(\bm{\Xi}^u(\bm{q}^u,\bm{m}^u)\big)  -\sum_u\sum_{a<b}\hat{q}^{ab} q^{ab}-\sum_{au} \hat{m}^{au} m^{au} \\
    .    
\end{align}
}Our ansatz will reflect the fact that each student possesses its own set of data: $q^u_{ab}=q^u$, $m^u_a=m^u$, across all replicas $a,b$. According to this we can rewrite the above free energy density as

\begin{align}
-\beta f(\bm{q}^u,\bm{m}^u;\bm{\hat{q}}^u,\bm{\hat{m}}^u) & \approx   \frac{1}{2} \sum_u\hat{q}^{u} q^{u} -\sum_u \frac{\hat q^u}{2} -\sum_{u} \hat{m}^{u} m^{u} \nonumber \\
    &-\sum_u \frac{\alpha_u}{2} \ln \left( (1-\hat\beta)(1-\beta+\beta q^u) \right) +\sum_u \frac{\alpha_u}{2} \frac{\beta (1-\hat\beta)q^u + \hat\beta \beta (m^u)^2}{(1-\hat\beta)(1-\beta+\beta q^u)} \nonumber \\
    &+ \ln \mathbb{E}_{\{s^u\}} \int \prod_u \mathcal{D}z^u \ln \sum_{\{\xi^u\}} \exp \left( \sum_{u} \xi^u h^u(z^u)+J \sum_{u<v}\xi_{u}\xi_{v}\right) \,,
\end{align}
where we defined: $h^u(z^u):= \hat m^u + z^u \sqrt{\hat q^u} $ and $J:= {\gamma}/{y}$. The extremal points are the ones with the order parameters satisfying the following equations
 \begin{align}
    \hat{m}^u= &\frac{\alpha_u \hat{\beta}\beta m}{(1-\hat{\beta})(1-\beta+\beta q^u)} \\
    \hat{q}^u =&  \frac{\alpha_u \hat{\beta} \beta^2(m^u)^2}{(1-\hat{\beta})(1-\beta+\beta q^u)^2}+\frac{\alpha_u \beta^2 q^u}{(1-\beta+\beta q^u)^2}
    \end{align}
    and
 \begin{align}
    m^u&=\int \prod_u Dz^u \sum_{\{\xi^u\}}P_J \left( \{\xi^u\} \right) \xi^u 
    \label{Aeq: SDD ms}\\ 
    q^u &= \int \prod_u Dz^u \left(\sum_{\{\xi^u\}}P_J \left( \{\xi^u\} \right) \xi^u \right)^2
    \label{Aeq: SDD qs}
\end{align}
with the probability $P_J \left( \{\xi^u\} \right)$ being
\begin{equation}
    P_J \left( \{\xi^u\} \right)  = z_J^{-1} \exp \left( \sum_{u} \xi^u h^u(z^u)+J \sum_{u<v}\xi_{u}\xi_{v}\right) \,. 
\end{equation}

\subsection{Effect of $\gamma$ on the \textit{P-sR} transition. Crossover from quadratic to linear constraints}
We consider the linearized self--consistency equations for the magnetizations~\(m^u\), obtained by expanding Eqs(\ref{Aeq: SDD ms}-\ref{Aeq: SDD qs}) around the paramagnetic
solution \(m^u,q^u\sim0\). In doing this we should be able to characterize the sR transition upon lowering the temperatures and see the connection between the coupling strength and the student loads. For each $u$-student, the linearized equations read
\begin{equation}
m^u
=
k_u\, m^u
+
\sum_{v\neq u} k_v\, \mathbb{E}_P[\xi^u \xi^v]\, m^v,
\label{eq:linearized}
\end{equation}
where
\begin{equation}
k_u
=
\frac{\alpha^u \hat\beta \beta}{\Delta_T (1-\beta)},
\end{equation}
and \(\mathbb{E}_P[\cdot]\) denotes expectation with respect to the fully
connected Ising measure
\begin{equation}
P(\boldsymbol{\xi})
\propto
\exp\!\left(
J
\sum_{u< v} \xi^u \xi^v
\right).
\label{Aeq: transition heter data measure}
\end{equation}
\\
For two students (\(u=1,2\)), symmetry implies
\begin{equation}
\mathbb{E}_P[\xi^1 \xi^2] = C,
\qquad
C = \tanh(J),
\end{equation}
where \(C\in[0,1]\) is a smooth, monotonically increasing function of \(\gamma\).
The linearized system becomes
\begin{align}
m^1 &= k_1 m^1 + C k_2 m^2, \\
m^2 &= k_2 m^2 + C k_1 m^1.
\end{align}
A non--trivial solution \((m^1,m^2)\neq(0,0)\) exists when the determinant of
the corresponding linear system vanishes:
\begin{equation}
(1-k_1)(1-k_2) - C^2 k_1 k_2 = 0,
\label{eq:det}
\end{equation}
or equivalently
\begin{equation}
(1-C^2) k_1 k_2 - k_1 - k_2 + 1 = 0.
\label{Aeq: quadratic}
\end{equation}

\begin{figure}[H]
    \centering
    
     
    \begin{minipage}{0.4\linewidth}
        \centering
        \includegraphics[width=0.9\linewidth]{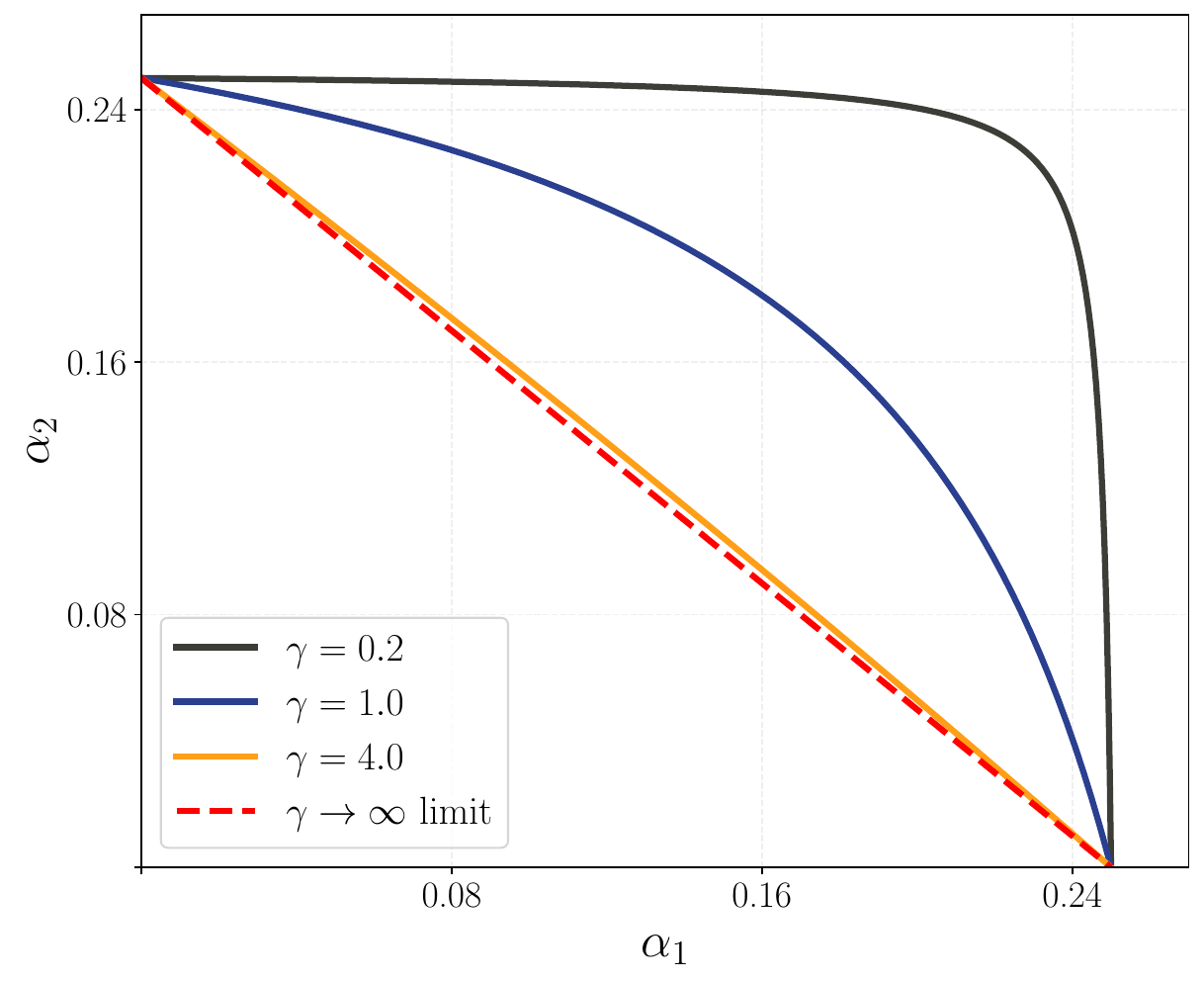}
    \end{minipage}%
    \hspace{0.05\linewidth} 
    \begin{minipage}{0.4\linewidth}
        \centering
        \includegraphics[width=\linewidth]{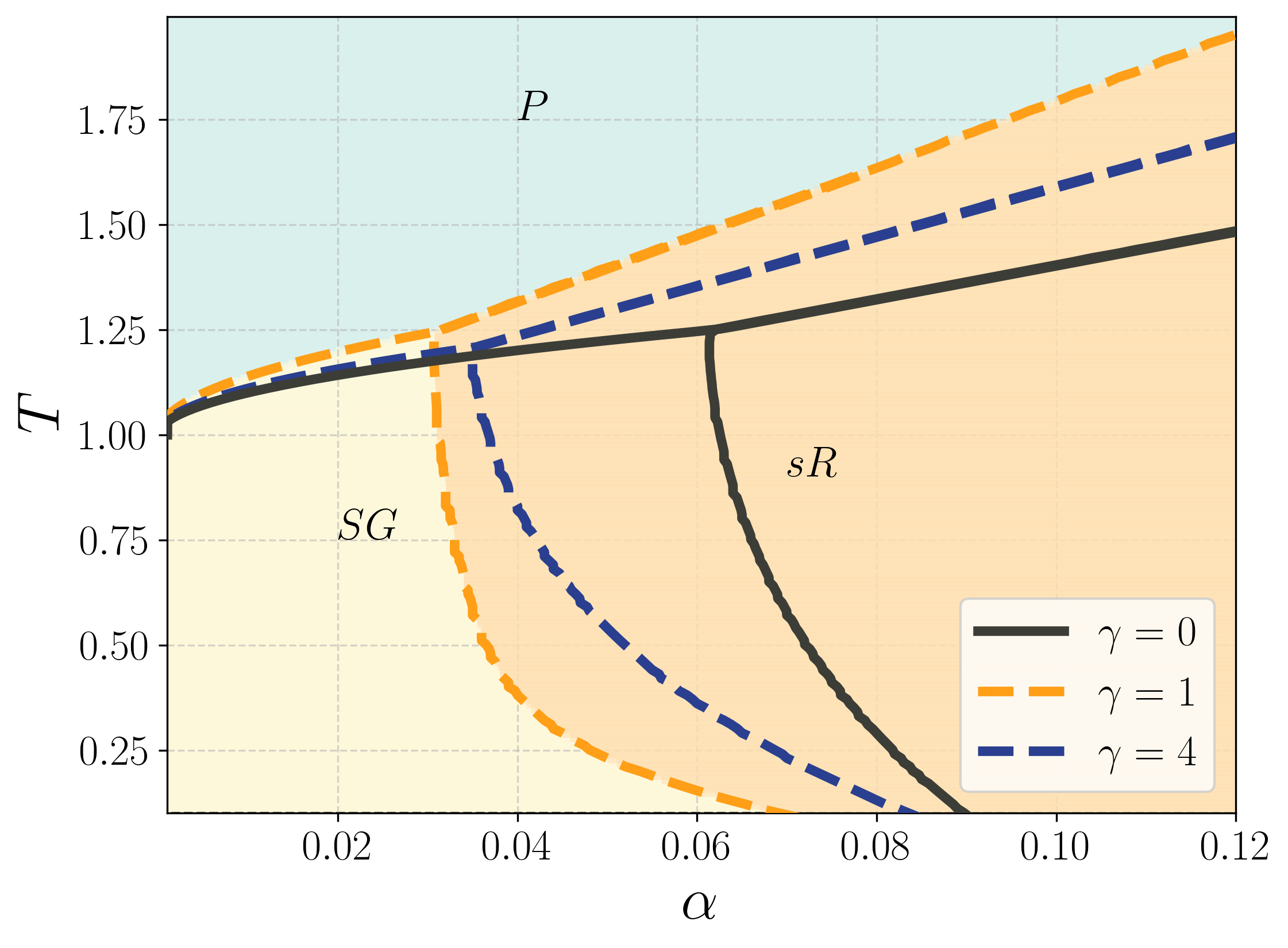}
    \end{minipage}
\caption{
\textbf{Left:}Dependence of the transition line according to different $\gamma$ values upon fixing $\beta=0.5$. At small cpoupling strenght the transition line defines an hyperbola. The more $\gamma$ gets larger the more $C(\gamma)$ saturates and the boundaries between the P and sR transition becomes a linear relation. \textbf{Right:} Phase diagram of the heterogeneous data regime for two interacting students for $\alpha_1=\alpha_2=\alpha$. We can appreciate the effect of $\gamma=0,1,4$ on the surface dividing the sR region from the other two phases.   
}
\label{Afig: transition vs gamma}
\end{figure}
For finite coupling, Eq.~\eqref{Aeq: quadratic} defines a quadratic
(hyperbolic) constraint in the positive quadrant
\(\alpha^1,\alpha^2>0\).
The physically admissible region, and so the one containing the transition line of fig.~\ref{Afig: transition vs gamma} (left), is the one inside the interval 
\begin{equation}
(\alpha_1,\alpha_2) \in \Bigg[ 0,\frac{\Delta_T (1-\beta)}{\hat\beta \beta (1-C^2))}  \Bigg) \times \Bigg[0,\frac{\Delta_T (1-\beta)}{\hat\beta \beta (1-C^2))} \Bigg)\,.
\end{equation}
Since \(C(\gamma)\) depends smoothly on \(\gamma\), the coefficient
\(1-C^2\) decreases continuously from \(1\) to \(0\). Consequently,
Eq.~\eqref{Aeq: quadratic} progressive saturate from a quadratic
(hyperbolic) relation at moderate coupling to the linear constraint
\begin{equation}
\label{Aeq: linear data split}
    \alpha_1+\alpha_2 = \frac{\Delta_T (1-\beta)}{\hat\beta \beta}
\end{equation}
 in the $y\to\infty$ coupling regime.
In the SDN case obtaining the solutions of the order parameter equations becomes harder the more students are involved, due to the $\mathcal{O}(y)$ number of gaussian integrations. However the effect of large coupling limit at the transition towards the magnetization (\ref{eq:linearized}) can be studied for arbitrary students. To do so we notice the sR-transition line solution is equivalent of asking the following matrix $M$ is singular:
\begin{equation}
M \bm m =\bm 0\,,
\end{equation}
where the entries of the matrix are $M_{uv}=\delta_{uv} d_u-C k_u$ and $d_u=1-(1-C)k_u$ and $C$ is now the correlation on the measure (\ref{Aeq: transition heter data measure}) for arbitrary 
number of students. Since  $D=\textbf{diag}(d_u)$ the singularity condition on $M$ can be obtained by imposing
\begin{equation}
0=\det M= \det (D-C \bm 1\bm k^T )
=\det D (1-C \bm k^T D^{-1} \bm 1)
\end{equation}
where now $\bm k =(k_1,k_2,...)$ and $\bm 1$ is the one vector of dimension $y$. The condition 
$1-C \bm k^T D^{-1} \bm 1=0$ can be rewritten explicitely as
\begin{equation}
\sum_{u=1}^{y} \frac{k_u}{1-(1-C),k_u} = \frac{1}{C}.
\label{Aeq: determinant transition condition Heter data}
\end{equation}
Finally we need to compute $C$. We use $\sum_{u\neq v}\xi^u \xi^v=\Big(\sum_u \xi^u\Big)^2-y$:
\begin{equation}
    C=\mathbb{E}_P[\xi^u \xi^v]=\frac{\sum_{\xi^u}e^{
\frac{J}{2}\Big(\sum_u \xi^u\Big)^2 
} \xi^u \xi^v}{\sum_{\xi^u}e^{
\frac{J}{2}\Big(\sum_u \xi^u\Big)^2 
}}=\frac{\int Dz\sum_{\xi^u}e^{
z\sqrt{J}\sum_u \xi^u 
} \xi^u \xi^v}{\int Dz\sum_{\xi^u}e^{
z\sqrt{J}\sum_u \xi^u 
}}=\frac{\int Dz \sinh^2 (z\sqrt{J})\cosh(z\sqrt{J})^{y-2}}{\int Dz \cosh(z\sqrt{J})^{y}}\,.
\end{equation}
 When $\gamma \to \infty$ we still have $C \to 1$ and Eq.(\ref{Aeq: determinant transition condition Heter data}) becomes
 \begin{equation}
     \sum_{u=1}^y k_u =1
 \end{equation}
      \begin{equation}    
      \sum_{u=1}^y \alpha_u =\frac{\Delta_T(1-\beta)}{\hat\beta \beta}
 \end{equation}
 meaning the students can split linearly the load of the WA non-corrupted case to recover the teacher. When data are , e.g. evenly distributed ($\alpha_1 = \alpha_2 = \alpha$), the system's behavior can be described using the same $(\alpha, T)$ phase diagram from the previous section. The right panel of Fig. \ref{Afig: transition vs gamma} illustrates how both the sR and SG transition lines shift with increasing $\gamma$. Similar to the SDN cases, the transition boundary moves toward lower loads and higher temperatures.
 
\section{Monte Carlo Simulations}
The system of interacting Hopfield models can be seen as an extended RBM as shown in Fig.~\ref{fig: structured hidden layer RBM}.
From the posterior in (\ref{eq: student posterior2}) we can decouple both the data and the interacting Hamiltonians and reveal a bipartite structure
\begin{equation}
\label{eq: hamiltonian structured RBM}
\mathcal{H} = -\sqrt{\frac{\beta}{N}}\sum_{u,\mu}\left[x_{u\mu}\sum_{i}\xi_{i}^{u}\eta_{i}^{u,\mu}\right]  -\sqrt{\frac{\gamma}{y}}\sum_{i}\left[z_{i}\sum_{u}\xi_{i}^{u}\right]\,,
\end{equation}

 with related measure 
\begin{align}
\hspace{-2em}P_{c}(\bm{\xi}|\bm{\mathcal{J}}) & ={Z}_{c}^{-1}(\bm{\mathcal{J}})\int\prod_{u,\mu}\mathcal{D}x_{u,\mu}\exp\left(\sqrt{\frac{\beta}{N}}\sum_{u,\mu}\left[x_{u,\mu}\sum_{i}\xi_{i}^{u}\eta_{i}^{u,\mu}\right]\right) \nonumber \\
 & \int\prod_{i}\mathcal{D}z_{i}\exp\left(\sqrt{\frac{\gamma}{y}}\sum_{i}\left[z_{i}\sum_{u}\xi_{i}^{u}\right]\right)\,. \label{eq: MixedBipartite}
\end{align}

\begin{figure}[H]
    \centering 
    \begin{minipage}{0.4\linewidth}
        \centering
        \includegraphics[width=\linewidth]{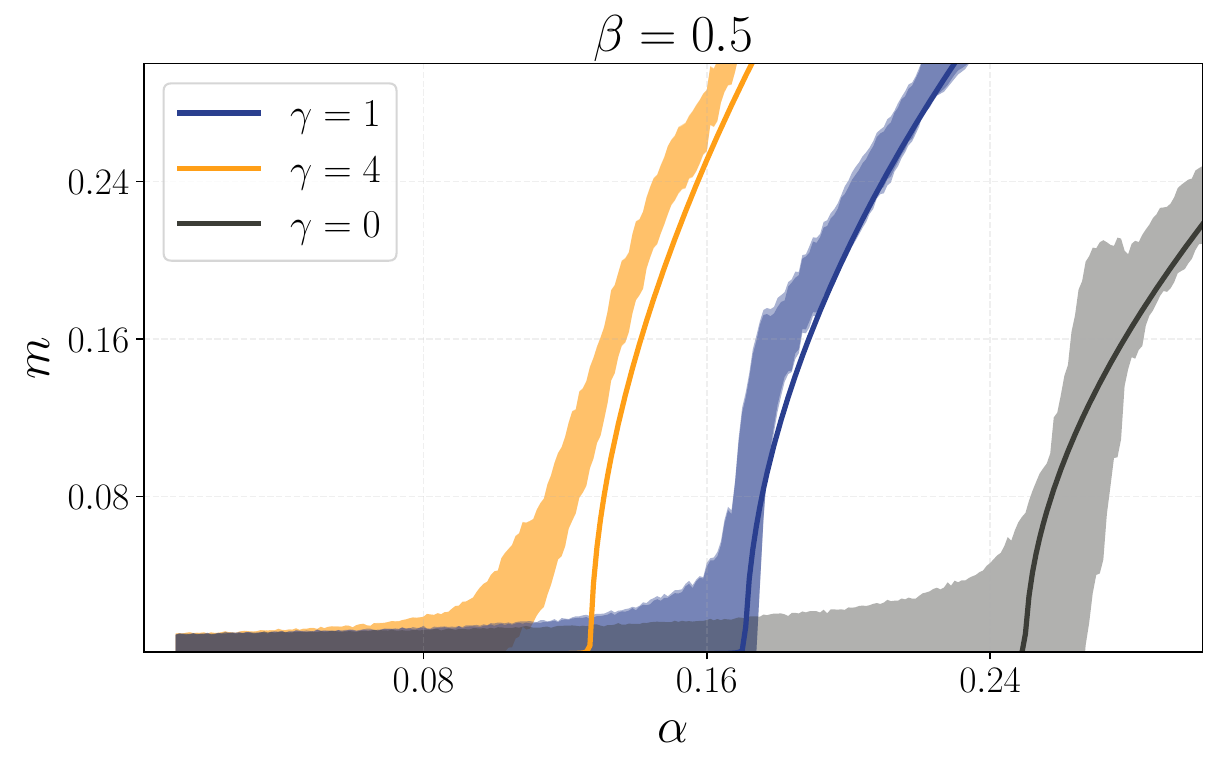}
    \end{minipage}%
    \hspace{0.05\linewidth} 
    \begin{minipage}{0.4\linewidth}
        \centering   \includegraphics[width=\linewidth]{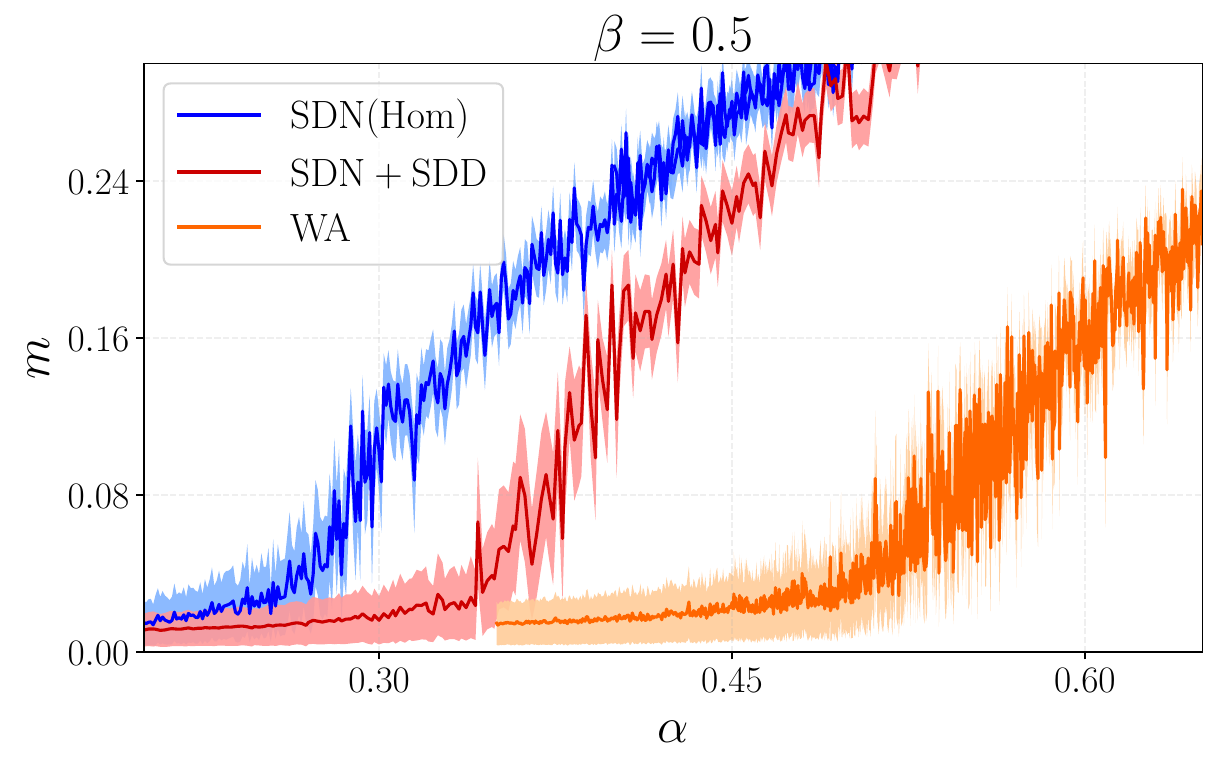}
    \end{minipage}
\caption{
MC simulations for a system of two interacting students facing a dataset with baseline noise $\hat{\beta}=0.8$.
\textbf{Left:} Results for the SDD version where the dataset is provided uncorrupted to the system. Magnetizations are plotted across different coupling strengths ($\gamma=0, 1, 4$) for the case $\alpha_1=\alpha_2 =\alpha$. Solid lines represent theoretical predictions. Increasing $\gamma$ reduces the critical load required to magnetize the system, dropping from $\alpha_{\mathrm{Hop}} \approx 0.24$ in the uncoupled ($\gamma=0$) case to $\alpha \approx 0.12$ ($\gamma=4$), which already approaches the linear saturation limit (\ref{Aeq: linear data split}).
\textbf{Right: }A comparison of three distinct cases: SDN-homogeneous, SDN+SDD, both with $\gamma=1$, and a WA student. For the first two variants, the corruption rate is set to $r=0.7$ to match the WA version. The combined FL scenario (SDN+SDD) introduces two distinct sources of error: data corruption and differing dataset realizations. Consequently, a larger amount of data is required to initiate teacher recovery compared to the SDN-homogeneous or the SDD ($r=1$) case on the left panel, though it still outperforms the standalone WA. }
\label{Afig: simulation ransition}
\end{figure}

The two sets of Gaussian variables ($z_i,x_i$) are now responsible to mediate the data and the students interacions and represent the bipartite version of the model.
The three types of units are related by a conditional relation that allow us to perform the following alternate sampling:
\begin{align}
\label{Aeq: sampling}
p(\bm{\xi} \mid \bm{x}, \bm{z}) &\propto \prod_{i,u} \exp\left\{ \xi_{i}^{u} \left[ \sqrt{\frac{\beta}{N}}\sum_{\mu}\eta_{i}^{u,\mu}x_{\mu,u} + \sqrt{\frac{\gamma}{y}}z_{i} \right] \right\} \,, \\
p(\bm{x} \mid \bm{\xi}) &\propto \prod_{\mu,u} \exp\left\{ -\frac{1}{2}x_{\mu,u}^2 + \sqrt{\frac{\beta}{N}} \, x_{\mu,u}\sum_{i}\xi_{i}^{u}\eta_{i}^{u,\mu} \right\} \,, \\
p(\bm{z} \mid \bm{\xi}) &\propto \prod_{i} \exp\left\{ -\frac{1}{2}z_i^2 + \sqrt{\frac{\gamma}{y}} \, z_{i}\sum_{u}\xi_{i}^{u} \right\} \,.
\label{Aeq: sampling z}
\end{align}
In the main text and in Fig.~\ref{Afig: simulation ransition} we perform the simulations of the model in both the (i) SDN and (ii) SDD scenarios. For (i) we generate the underlined data that will be successively corrupted with same or different noise level, to simulate both homogeneous and heterogeneous ansatz. The base dataset is extracted by simulating a Hopfield model at temperature $\hat\beta=0.8$, with one pattern $\hat{\bm \xi}=\bm 1$, with sytem size $N=10000$ spins and generate a number of $M$ datapoints in order to reproduce the load $\alpha=M/N$. The entire dataset is further copied as much as the desired students involved, in our case $y=2$. As described in the main text each sample inside the respective dataset is then radomly flipped, according to the corruption rates $r_1,r_2$ of (\ref{Aeq: corruption distribution}). Subsequently we follow PCD prescription to reach the equilibrium distribution of the extended RBM. This is done performing one sampling step which consist in updating each units of the bipartite network, starting from a random configuration $\bm \xi^0$ and following (\ref{Aeq: sampling}-\ref{Aeq: sampling z}). We update the persistent chain for $2\times10^3$ times for a number of $N_s=2\times10^3$ samples. 

A second type of simulation is performed for case (ii). The main difference w.r.t. (i) lies in the teacher samples. In this case we generate a specific realization of $M^u=\alpha^u N$ samples, one for each student ($u=1,2$) and use these as patterns for the RBM without introducing the corruption randomness. All the other specifics remain the same as before.

In Fig.~\ref{Afig: simulation ransition} (left) we see the effect of different coupling values $\gamma=\{0,1,4\}$ on the magnetization of the SDD scenario $y=2$, without the presence of any corruption (here $\bm r=\bm1$). As predicted by Eq.(\ref{Aeq: quadratic}) when we increase $\gamma$ we move the P-sR transition to a lower load and the more we increase $\gamma$ the more the transition point get close to the linear splitting (\ref{Aeq: linear data split}) $\alpha=\alpha_{\mathrm{Hop}}/2$, where $\alpha_{\mathrm{Hop}} \sim 0.24$ correspond to the single Hopfiled model transition when $\beta=0.5$ \cite{alemanno2023hopfield}.
By leaveraging the simulations, it is also possible to investigate the mixed SDN+SDD case. To achieve this, we first realize two different instances of datasets from the teacher, as discussed above for (ii).  After a successive random flip, with rates $r_1, \ r_2$, we sample this mixed system which magnetization is shown in Fig.~\ref{Afig: simulation ransition} (right). The image shows the SDN-homogeneus at $r=0.7$ (blue) compared to its WA, $r=0.7$ version (orange) and the SDN+SDD (red). In this last case we can see the transition happen after the SDN one, but still before the WA. In this case the energy landscape of the system is more involved due to the combination of structural and corruption noise. This make the system more difficult to navigate towards the teacher signal w.r.t. the pure SDN version. 

\end{document}